\documentclass[]{spie}  
\usepackage[]{graphicx}
\usepackage{amsmath,amssymb}
\usepackage{subcaption}
\usepackage{multirow}
\usepackage{url}

\def\mf{\mathbf}

\newcommand{\reffig}[1]{Figure \ref{#1}}
\newcommand{\refsec}[1]{Section \ref{#1}}
\newcommand{\refnum}[1]{Ref.~\citenum{#1}}
\DeclareMathOperator{\rank}{rank}

\title{Mining the GPIES database}

\author{Dmitry Savransky\supit{a,b}, Jacob Shapiro\supit{a}, Vanessa Bailey\supit{c}, Robert De Rosa\supit{d}, Jason Wang\supit{d}, Jean-Baptiste Ruffio\supit{e}, Eric Nielsen\supit{e}, Melisa Tallis\supit{e}, Marshall Perrin\supit{f}
 \skiplinehalf
\supit{a}Sibley School of Mechanical and Aerospace Engineering,\\ Cornell University, Ithaca, NY, 14853, USA\\
\supit{b}Carl Sagan Institute, Cornell University, Ithaca, NY, 14853, USA\\
\supit{c}Jet Propulsion Laboratory, California Institute of Technology, Pasadena, CA, 91109, USA\\
\supit{d}University of California at Berkeley, Berkeley, CA, 94720, USA\\
\supit{e}Kavli Institute for Particle Astrophysics and Cosmology, Stanford University,\\ Stanford, CA, 94305, USA\\
\supit{f}Space Telescope Science Institute, Baltimore, MD, 21218, USA
}

\authorinfo{Send correspondence to Dmitry Savransky 
\texttt{ds264@cornell.edu}}

\begin{document} 
\maketitle 

\begin{abstract}
The Gemini Planet Imager Exoplanet Survey (GPIES) is a direct imaging campaign designed to search for new, young, self-luminous, giant exoplanet.  To date, GPIES has observed nearly 500 targets, and generated over 30,000 individual exposures using its integral field spectrograph (IFS) instrument.  The GPIES team has developed a campaign data system that includes a database incorporating all of the metadata collected along with all individual raw data products, including environmental conditions and instrument performance metrics.  In addition to the raw data, the same database also indexes metadata associated with multiple levels of reduced data products, including contrast measures for individual images and combined image sequences, which serve as the primary metric of performance for the final science products.  Finally, the database is used to track telemetry products from the GPI adaptive optics (AO) subsystem, and associate these with corresponding IFS data.

Here, we discuss several data exploration and visualization projects enabled by the GPIES database.  Of particular interest are any correlations between instrument performance (final contrast) and environmental or operating conditions. We show single and multiple-parameter fits of single-image and observing sequence contrast as functions of various seeing measures, and discuss automated outlier rejection and other fitting concerns.  We also explore unsupervised learning techniques, and self-organizing maps, in particular, in order to produce low-dimensional mappings of the full metadata space, in order to provide new insights on how instrument performance may correlate with various factors.  Supervised learning techniques are then employed in order to partition the space of raw (single image) to final (full sequence) contrast in order to better predict the value of the final data set from the first few completed observations. Finally, we discuss the particular features of the database design that aid in performing these analyses, and suggest potential future upgrades and refinements.

\end{abstract}

\section{INTRODUCTION} \label{sec:intro} 

The Gemini Planet Imager Exoplanet Survey (GPIES)\cite{mcbride2011experimental,macintosh2014first} is a direct imaging campaign, started in 2014, with the goal of observing young, nearby main sequence stars to discover new, self-luminous, giant exoplanets.  To date, the survey has used approximately 700 of its allocated 895 hours of observing time, and has collected data on 484 targets, of a nominal list of 600 stars. The survey has observed multiple previously known exoplanets, has discovered a Jovian-analog exoplanet, 51 Eri b\cite{macintosh2015discovery}, and has generated numerous brown dwarf and circumstellar disk discoveries.

GPI's science data products are raw, 2D frames from the integral field spectrograph (IFS), which are packaged and stored in FITS format\cite{greisen2006representations} with multiple levels of header data encoding both the instrument, observatory, and ambient environment states at the time of the observation, along with target identification information. The raw frames are processed into reduced data products---3D spatial-spectral or spatial-polarimetric data cubes---via the GPI data reduction pipeline, described in depth in Refs.~\citenum{perrin2014gemini,perrin2016gemini,perrin2017gemini}. Full observing sequences of multiple individual observations are further reduced into second level products via a variety of post-processing algorithms including KLIP\cite{soummer2012detection,wang2015pyklip} and TLOCI\cite{marois2013tloci}.  KLIP-processed sequences are further analyzed via a forward model matched filter\cite{ruffio2017improving} based on the KLIP forward model described in \refnum{pueyo2016detection}, to automatically identify new point-source candidates in the data.  At each processing level, from the initial reduction through final matched filtering, the effective contrast (typically reported as the variation at the 5$\sigma$ level) serves as the primary metric of performance and is recorded as a function of angular separation and wavelength in individual images. Individual observations are reduced twice---first on the summit to create a `quicklook' image for on-site (or remote) observers to track instrument performance in real time, and again, with additional processing steps, prior to the reduction of full observing sequences.  The quicklooks encode average contrasts over the full wavelength band at three fiducial separations---0.25, 0.4 and 0.8 arcseconds---and store these in the reduced image header along with all other meta data. In the science grade processing, contrast profiles are generated as ancillary products, encoding the radially averaged contrasts between the inner and outer working angles of GPI's coronagraph for each distinct wavelength slice generated by the data reduction pipeline (typically 37 per instrument wavelength band). Both angular\cite{marois2006angular} and spectral\cite{racine1999speckle} planet signal diversity are used in the post-processing of full observing sequences, producing angular-separation, wavelength and spectral model dependent contrast curves for the full sequence, which are stored as ancillary data products.

The entirety of the processing chain described above is executed automatically via the GPIES campaign data system, described in depth in \refnum{wang2018automated}.  Briefly, all GPI IFS data taken under the GPIES campaign program, partner programs, or as engineering or calibration data, is continuously synced from the Gemini South dataflow to a server at Cornell University, which runs the GPIES database, described in detail in \refsec{sec:db}.  All metadata is extracted from the raw and quicklook images and indexed in the database, and the data itself is synced to the broader GPIES team via Dropbox.  Automated processing facilities at UC Berkeley, HIA, Stanford, and elsewhere pick up the raw data and created science grade reduced data products and level two reduced products which are subsequently synced via Dropbox and are also ingested in to the database. A variety of frontend tools interact with the database backend to provide analysis and planning utilities to the GPIES team. GPI AO system telemetry is similarly synced to Dropbox and indexed in additional, cross-linked tables in the database.

Significant prior work has been done to characterize GPI AO performance using the IFS data\cite{poyneer2016performance,bailey2016status} and to characterize overall GPI performance variations as a function of operating conditions\cite{tallis2017air}.  Here, we are instead focused only on data-driven analyses, which do not require specific physical modeling or a deep understanding of the relevant systems.  This paper explores the types of data analyses that are enabled by the simple access to correlated science, engineering and environmental data enabled by the GPIES database, and presents a series of results with predictive power for GPIES contrasts given information that would be available to observers before, or at the immediate start, or an observing sequence.

\section{THE GPIES DATABASE}\label{sec:db}
\begin{figure}[ht]
\centering
\includegraphics[width=\textwidth]{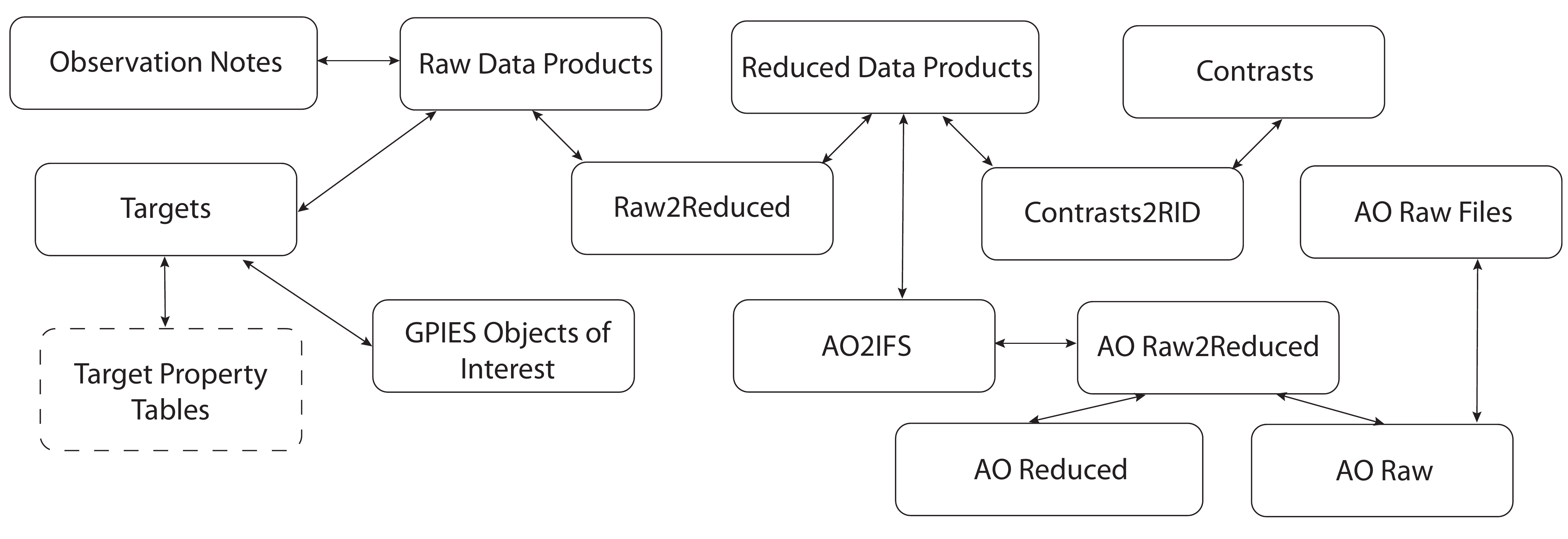}
\caption{Schematic of the GPIES database table structure. Arrows represent shared or referenced table indexes.\label{fig:db}}
\end{figure}
The GPIES database is implemented in structured query language (SQL) using the MySQL open source relational database management system. The database, shown schematically in \reffig{fig:db}, is implemented as multiple, cross-linked many-to-many relationships, with each relationship represented by a minimum of three tables.  Auxiliary cross-linked tables store additional data used for planning purposes, and the database also links to the full GPIES target database developed by Inseok Song at the University of Georgia.

The two primary many-to-many relationships are the raw and reduced IFS data products, and the raw and reduced AO telemetry products, which also link to any temporally correlated IFS products (continuous, full AO telemetry requires enormous data volumes not currently available at the Gemini South summit, and so telemetry data is taken periodically during GPIES observing runs and engineering testing, and referenced to science data in post-analysis).  The Contrasts database table stores contrast values from all science-grade reduced data products, including individual image and observing sequence reductions (quicklook contrasts are only stored for the three fiducial separations in the Reduced Data Product table). The Raw Data Products table has individual columns for every FITS keyword header key included in the raw IFS data, and the Reduced Data Product table has columns for all non-overlapping keywords (with the raw files) generated by the standard GPI data reduction pipeline and the team's internal post-processing codes.  

Additional keywords that may be present in custom reductions are stored in a single extra column blob in the Reduced Data Product table. Finally, an Observation Notes table stores binary flags (set at the summit during observations, or after the fact via a web interface) that mark bad raw data that should be excluded form observing sequence reductions, as well as any observer comments describing the specific data faults. This table is queried by the automatic reduction codes that generate the level two reduced data products.  The database currently contains approximately 14 GB of metadata and ancillary data products, indexing over 6 TB of data.  There are 136,151 indexed IFS raw data files, of which 30,142 are directly related to GPIES while the rest are engineering, calibration, and partner program data.  There are 263,973 reduced IFS data products, 86,092 raw AO telemetry data files, and 86,325,374 indexed contrast values.

\section{CONTRAST CORRELATIONS}\label{sec:correlations}

The first step in attempting to identify relationships with predictive power in the available data is to explore any correlations between the various types of metadata and the metric of interest, which in this case is the total observing sequence contrast.  We explore three parametric and nonparametric correlation tests:
\begin{enumerate}
\item The Pearson product-moment, defined as the covariance of two sample random variables, $\bar{x}, \bar{y}$, scaled by the product of their sample standard deviations:
\begin{equation}
r_{\bar x, \bar y} = \frac{E[(\bar x - \mu(\bar x))(\bar y - \mu(\bar y))]}{\sigma(\bar x) \sigma(\bar y)} \,.
\end{equation}
This metric identifies linear relationships between variables. 
\item The Spearman rank correlation, a non-parametric test for rank correlation, defined simply as the Pearson product moment of the ranks of the variables:
\begin{equation}
\rho_{\bar x, \bar y}  = r_{\rank \bar x, \rank \bar y} \,.
\end{equation}
\item The Kendall rank correlation, another non-parametric test for ordinal association, defined as the scaled difference between the number of concordant and discordant pairs in the sample data set:
\begin{equation}
\begin{split}
\tau = \frac{2}{n(n-1)}&\left(\sum_{i\ne j} \left[ ((x_i  > x_j ) \& (y_i > y_j)) | ((x_i  < x_j ) \& (y_i < y_j))\right]   \right.\\
&\quad {} - \left. \sum_{i\ne j} \left[ ((x_i  < x_j ) \& (y_i > y_j)) | ((x_i  > x_j ) \& (y_i < y_j))\right]   \right) \,.
\end{split}
\end{equation}
\end{enumerate}

\begin{figure}[ht]
\centering
\includegraphics[width=\textwidth,clip=true,trim = 1.5in 1.2in 1.5in 0.25in]{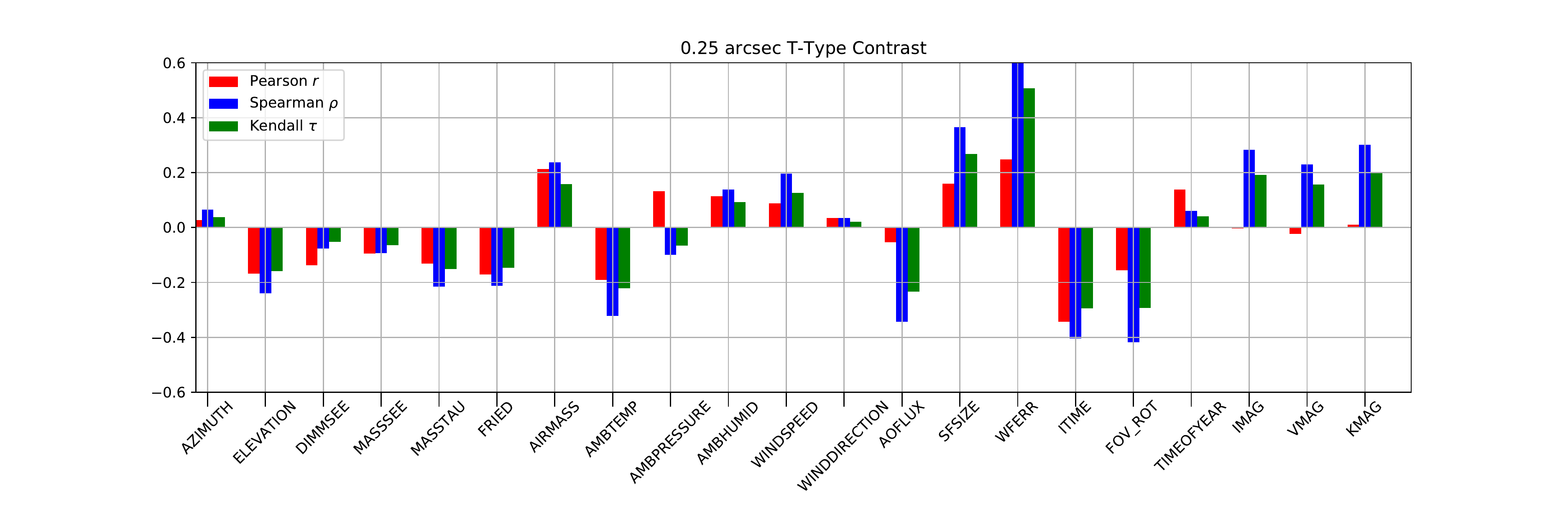}
\includegraphics[width=\textwidth,clip=true,trim = 1.5in 0.25in 1.5in 0.25in]{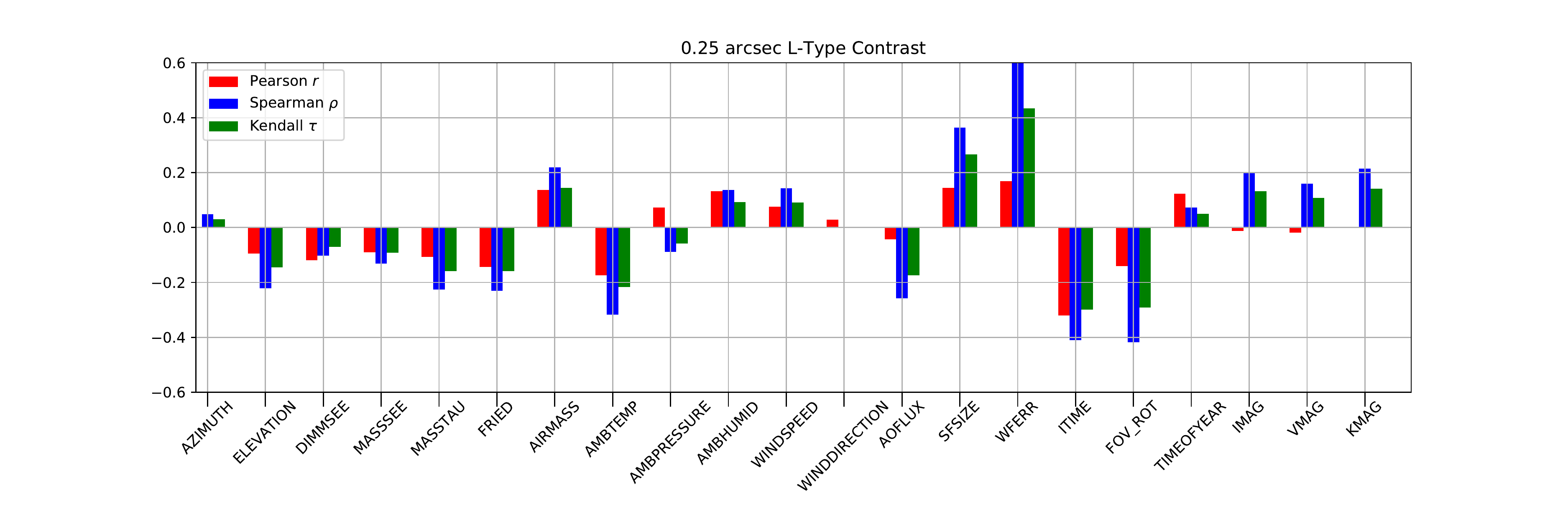}
\caption{Correlations of target, environmental, and instrument metadata with observing sequence contrast at 0.25 arcseconds, assuming either a T-type (\emph{top}) or an L-type (\emph{bottom}) planet spectrum in the reduction.\label{fig:corr025}}
\end{figure}

\begin{figure}[ht]
\centering
\includegraphics[width=\textwidth,clip=true,trim = 1.5in 1.2in 1.5in 0.25in]{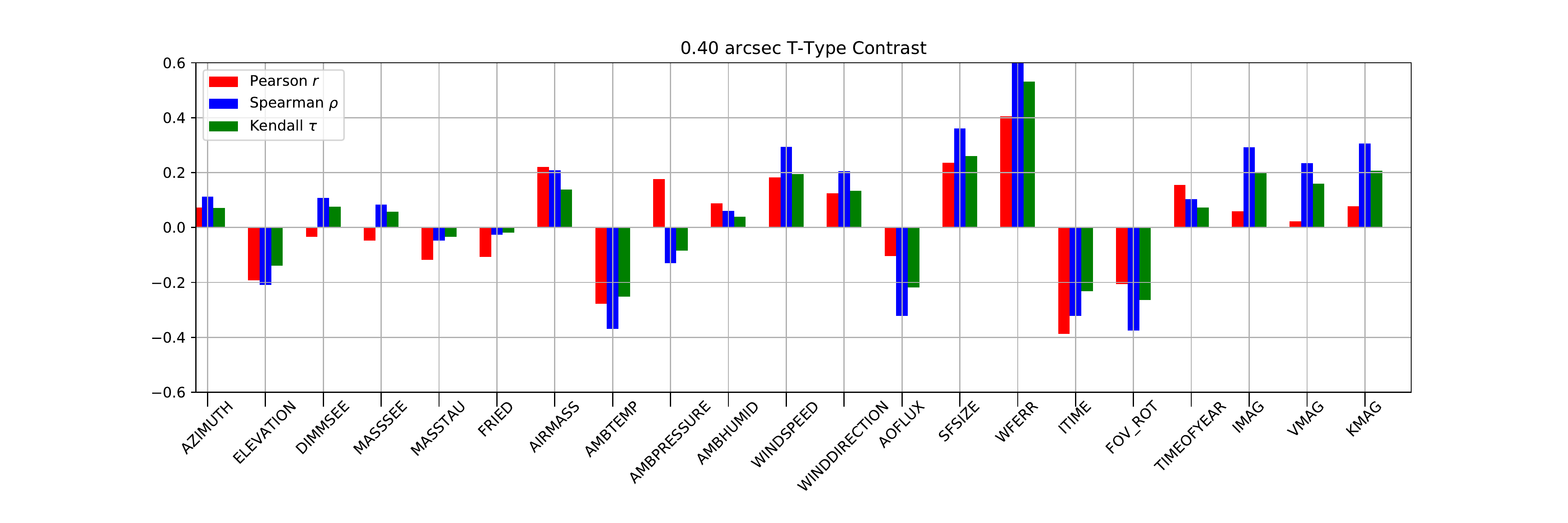}
\includegraphics[width=\textwidth,clip=true,trim = 1.5in 0.25in 1.5in 0.25in]{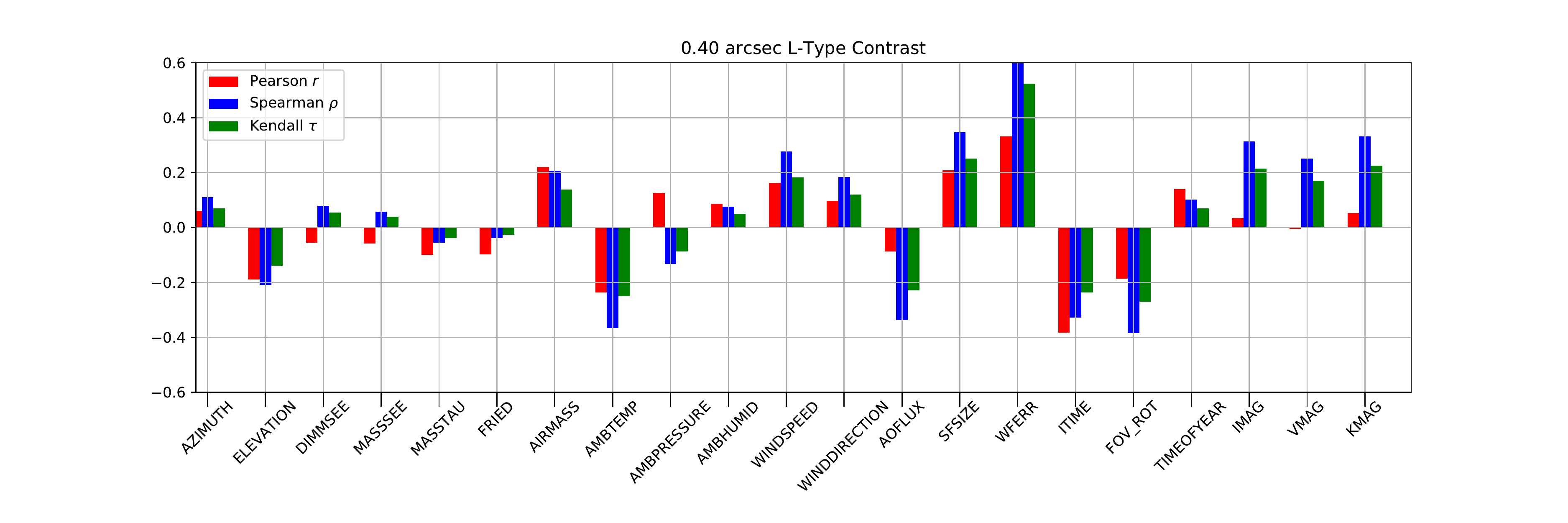}
\caption{Correlations of target, environmental, and instrument metadata with observing sequence contrast at 0.4 arcseconds, assuming either a T-type (\emph{top}) or an L-type (\emph{bottom}) planet spectrum in the reduction.\label{fig:corr040}}
\end{figure}

\begin{figure}[ht]
\centering
\includegraphics[width=\textwidth,clip=true,trim = 1.5in 1.2in 1.5in 0.25in]{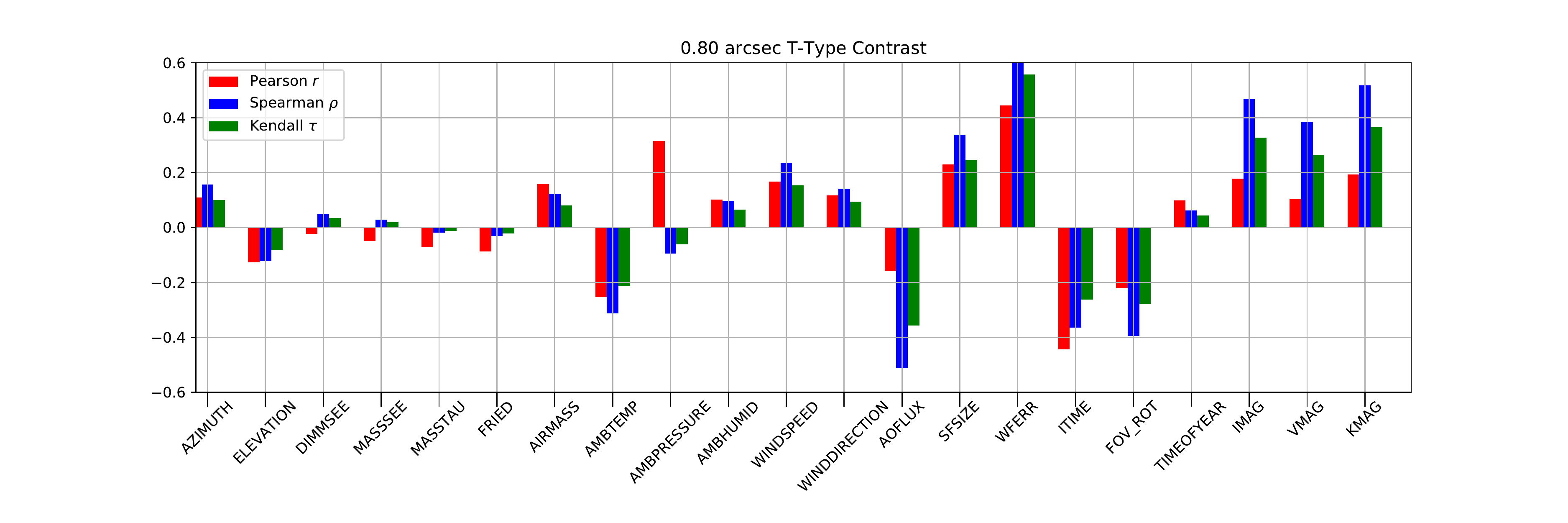}
\includegraphics[width=\textwidth,clip=true,trim = 1.5in 0.25in 1.5in 0.25in]{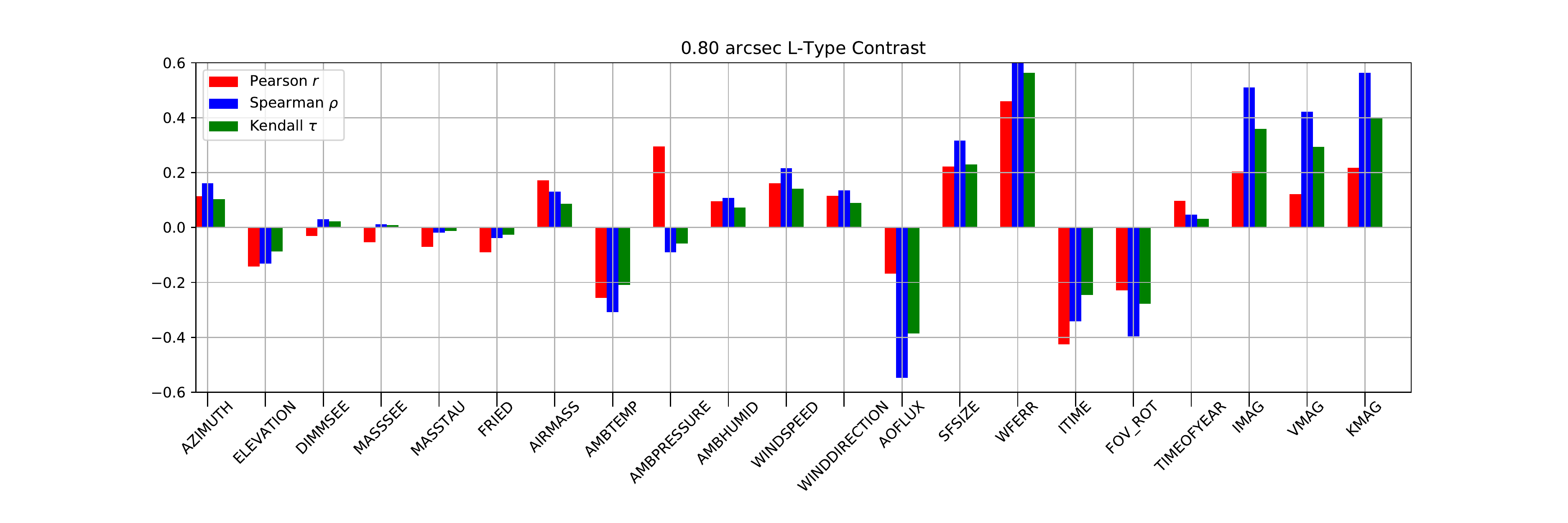}
\caption{Correlations of target, environmental, and instrument metadata with observing sequence contrast at 0.8 arcseconds, assuming either a T-type (\emph{top}) or an L-type (\emph{bottom}) planet spectrum in the reduction.\label{fig:corr080}}
\end{figure}

Figures \ref{fig:corr025} - \ref{fig:corr080} show the values of the three correlation coefficients against a variety of metadata sources for $\log_{10}$ contrasts evaluated at 0.25, 0.40 and 0.80 arcseconds, respectively, for 633 full GPIES observing sequences in the H band (the default band used for new targets), representing regular GPIES planet search or confirmation observations.  In each case, contrasts are evaluated assuming either a T or L spectral type spectrum for a potential planet (see, e.g., \refnum{nielsen2017evidence} for detailed discussions on contrast calculations for KLIP-processed GPI data).  The contrasts at particular angular separations are evaluated via a cubic interpolant of the full contrast curves to smooth out any anomalous spikes in the contrast curves due to numerical issues in the post-processing.  All three metrics are strictly defined between -1 and 1, with the sign representing a negative or positive correlation.   There are clear trends in the various factors with angular separation, confirming that noise is dominated by various factors at various points in the image plane. Most interestingly, however, is the fact that the correlations with the two contrast types do not scale together at the different angular separations, meaning that the choice of model in calculating the contrast impacts any performance model that can be fit to the contrasts.

The various values being correlated against can be grouped into three categories: target star properties, ambient environmental conditions during the observing sequence, and instrument settings and performance during the observing sequence.  The first category includes the star coordinates (encoded as azimuth and elevation) and magnitudes in the I, V, and K bands.  The TIMEOFYEAR value encodes the day number of the observation, and although it is not a target property, it is known before the start of an observation and therefore can be grouped with the static target properties.  The ambient conditions include seeing and $\tau_0$ estimates from the Gemini south MASS and DIMM monitors, along with the local airmass, the ambient temperature, humidity  and pressure, and the wind speed and direction.  The final category includes the size setting of the spatial filter in the AO system (see \refnum{poyneer2004spatially}), which is a proxy for how well the instrument is performing in response to local conditions, and the average wavefront error recorded by the AO system.  The final two values, ITIME and FOV\_ROT, encode the total integration time and the total field rotation over the observing sequence.  These values represent things that can be known well in advance of an observation, things that can be estimated shortly after the start of an observation, and things that can only be known at the end of the full observing sequence. 

\begin{figure}[ht]
\centering
\begin{subfigure}[c]{0.495\textwidth}
\includegraphics[width=\columnwidth,clip=true,trim = 0.1in 0.1in 0.5in 0.25in]{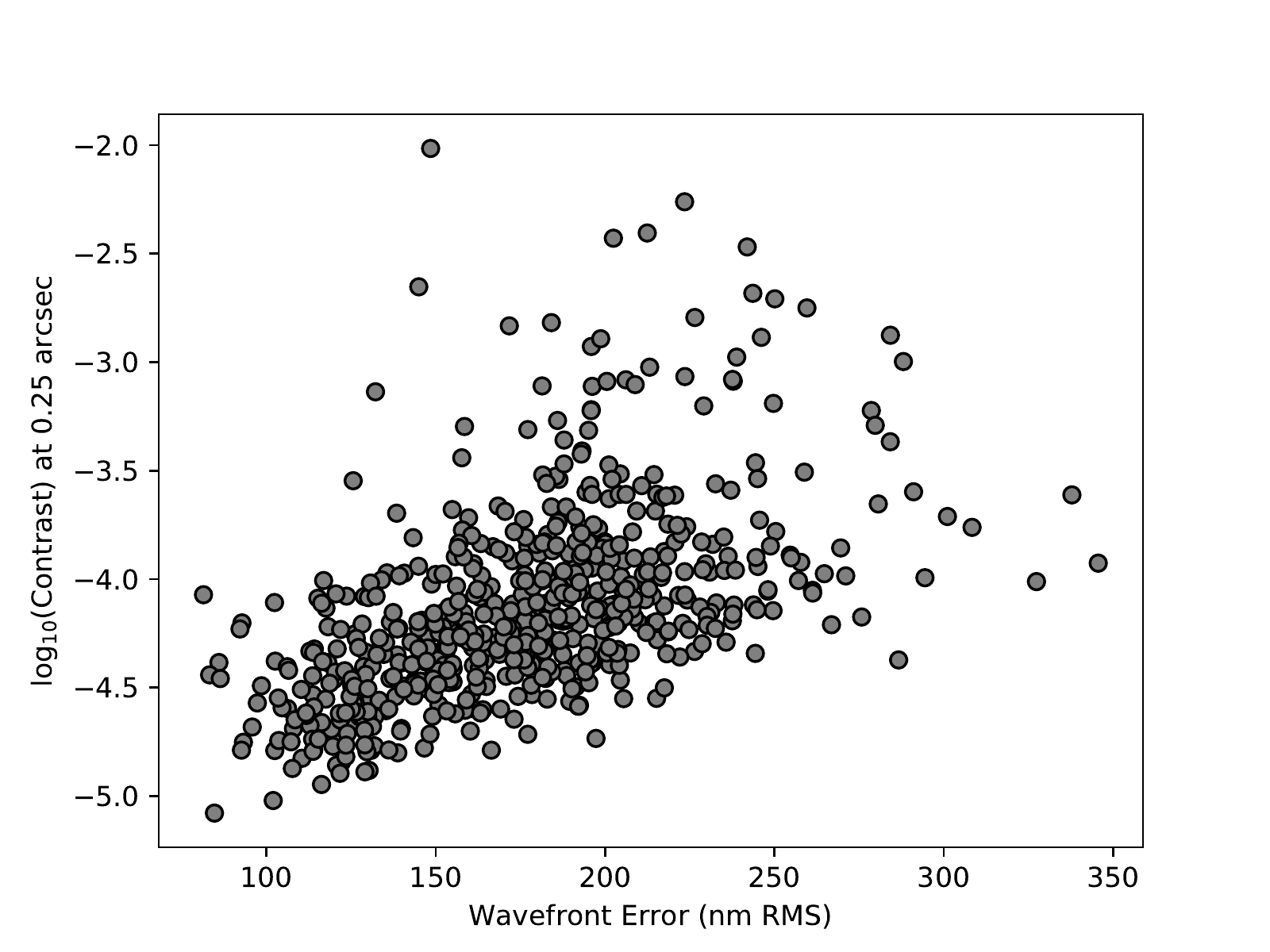}
\caption[]{Wavefront Error}
\end{subfigure}
\begin{subfigure}[c]{0.495\textwidth}
\includegraphics[width=\columnwidth,clip=true,trim = 0.1in 0.1in 0.5in 0.25in]{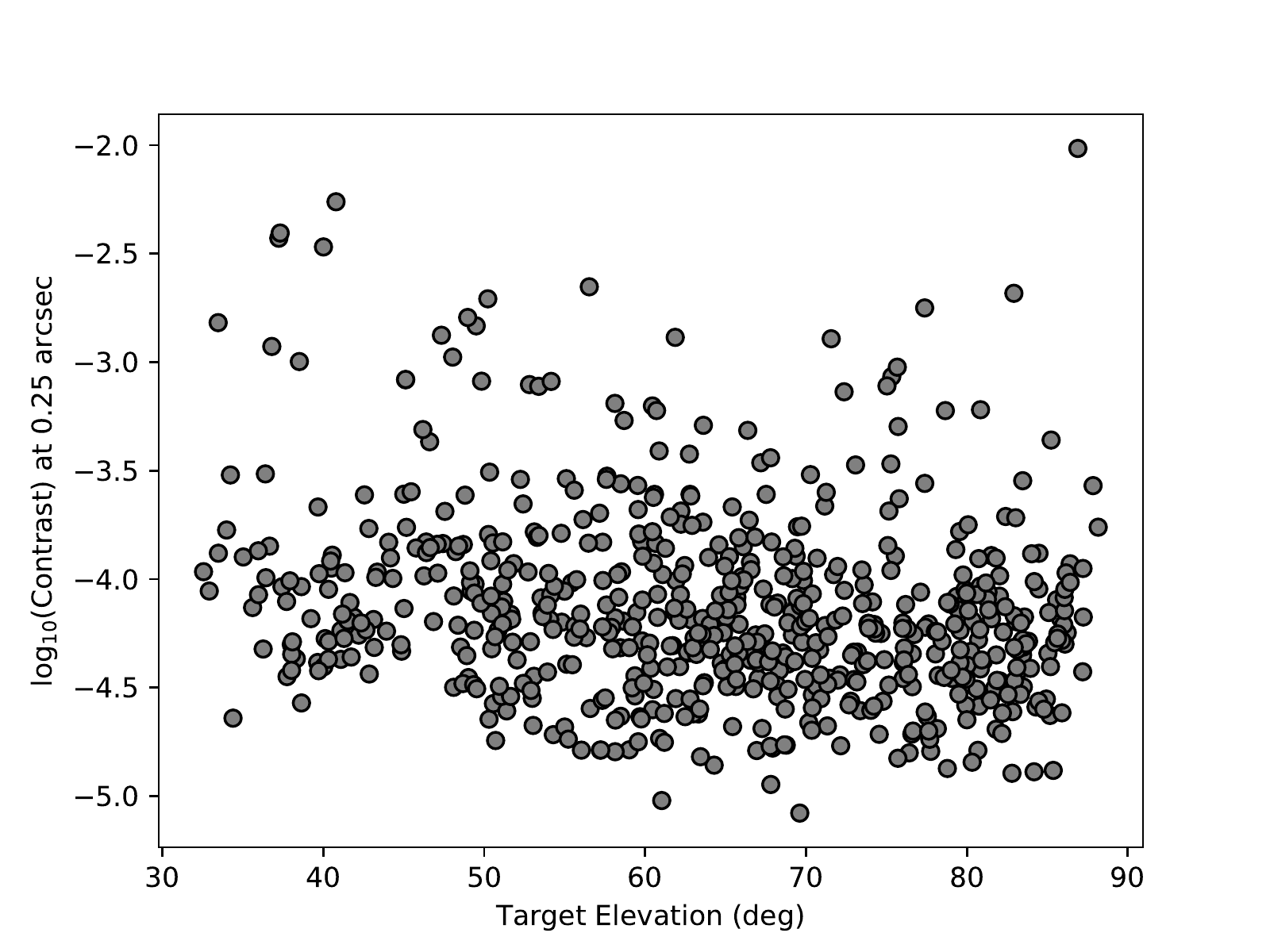}
\caption[]{Elevation}
\end{subfigure}
\caption{Scatter plots of $\log_{10}$ contrast for full observing sequences evaluated at 0.25 arcseconds angular separation as a function of wavefront error and elevation.\label{fig:scatter}}
\end{figure}

\reffig{fig:scatter} shows the scatter of the contrast at 0.25 arcseconds for the full observing sequences from GPIES in H band as a function of the wavefront error (highly positively correlated) and target elevation (moderately negatively correlated). While some linear relationships with the appropriate slopes may be evident in the data via visual inspection, we see also that the data is noisy, with significant spread about any linear model.  The root mean square error (RMSE) of linear fits for any of the parameters included in the correlation analysis have values of well over 0.5 $\log_{10}$ contrast, making them less than ideal for predictive purposes. 

\section{CONTRAST MODELING AND OUTLIER DETECTION}\label{sec:models}

To account for the inherent noisiness of the data, and the probable biasing of some data points by factors not captured in the metadata, we can try to partition the data set into a representative (foreground) set mixed with an outlier (background) set.  To do so, we follow the Bayesian methodologies laid out in Refs.~\citenum{hogg2010data,hogg2017data} and assume that the foreground data distribution is explainable by our data model $f_{\boldsymbol{\theta}}$, with parameters $\boldsymbol{\theta}$ and Gaussian noise, and the outliers all belong to a separate normal distribution with mean $\mu_o$ and standard deviation $\sigma_o$.  Whether a given data point $y_i$ belongs to the foreground or background distribution is determined by a binary random variable $o_i$.  Thus, the conditional probability of $y_i$, given the data $\mf x_i$ and model parameters is:
\begin{equation}
p(y_i|\mf x_i,\sigma_i,\boldsymbol{\theta},o_i,\mu_o,\sigma_o) = 
\frac{1}{\sqrt{2\pi\left(\sigma_i^2 + o_i\sigma_o^2\right)}}\exp \left(-\frac{[y_i - (1-o_i)f_{\boldsymbol{\theta}} (\mf x_i) - o_i \mu_o]^2}{2\left(\sigma_i^2 + o_i\sigma_o^2\right)}\right) \,.
\end{equation}
The marginalized likelihood for the full data set is therefore:
\begin{equation}
p(\{y_i\}_{i=1}^n|\{\mf x_i\}_{i=1}^n,\{\sigma_i\}_{i=1}^n,\boldsymbol{\theta},\mu_o,\sigma_o)= \prod_{i=1}^{n} \left [ Op(y_i|\mf x_i,\sigma_i,\boldsymbol{\theta},o_i=0) + (1-O) p(y_i|\mf x_i,\sigma_i,\boldsymbol\theta,o_i=1) \right] \,,
\end{equation}
where $O$ is a prior on $o$, selected to be:
\begin{equation}
p(o_i) = \begin{cases} O & o_i = 0 \\ 1 - O & o_i = 1\end{cases} \,.
\end{equation}

\begin{figure}[ht]
\centering
\begin{subfigure}[c]{0.495\textwidth}
\includegraphics[width=\columnwidth,clip=true,trim = 0.1in 0.1in 0.5in 0.25in]{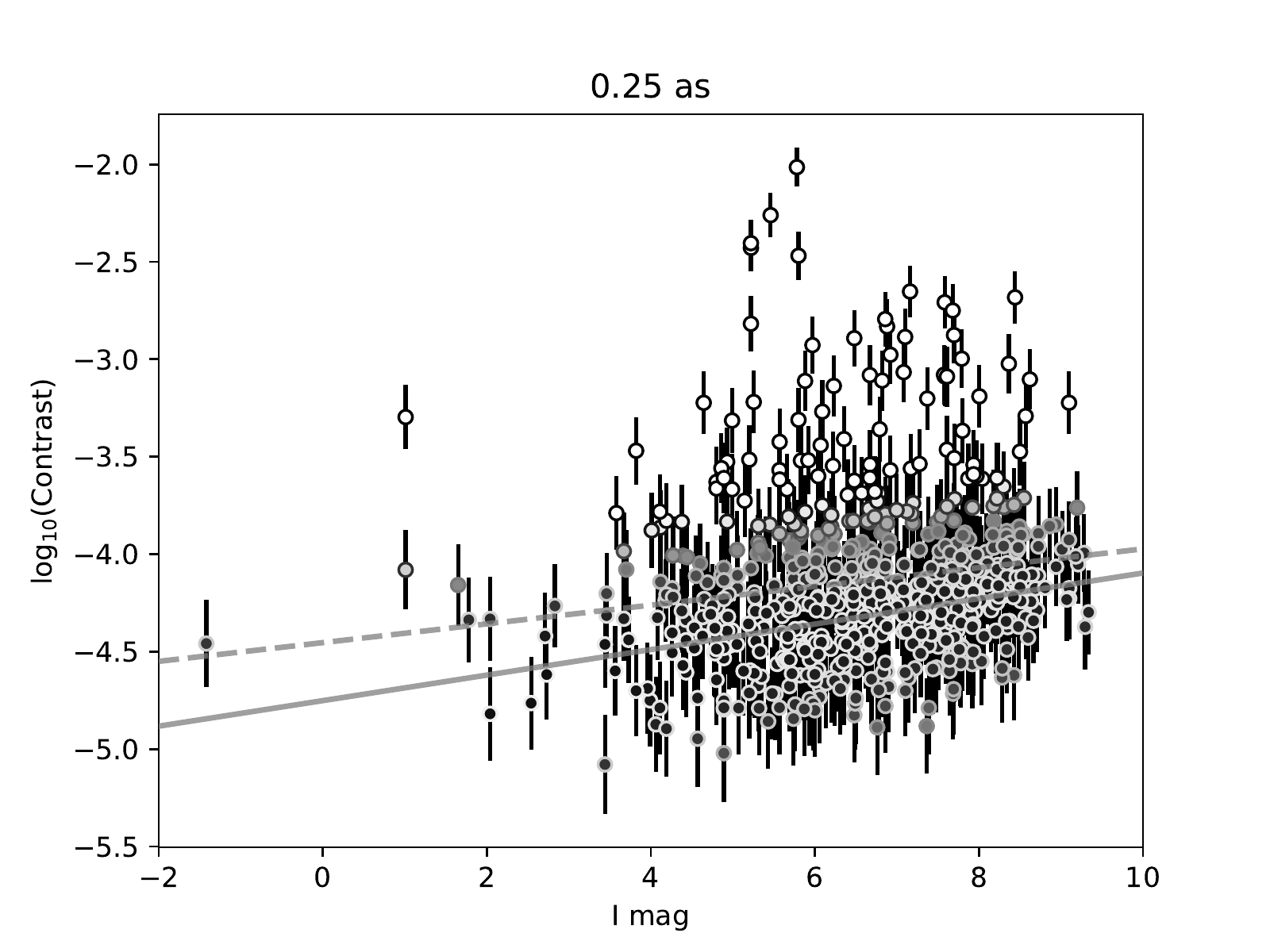}
\caption[]{0.25 arcseconds}
\end{subfigure}
\begin{subfigure}[c]{0.495\textwidth}
\includegraphics[width=\columnwidth,clip=true,trim = 0.1in 0.1in 0.5in 0.25in]{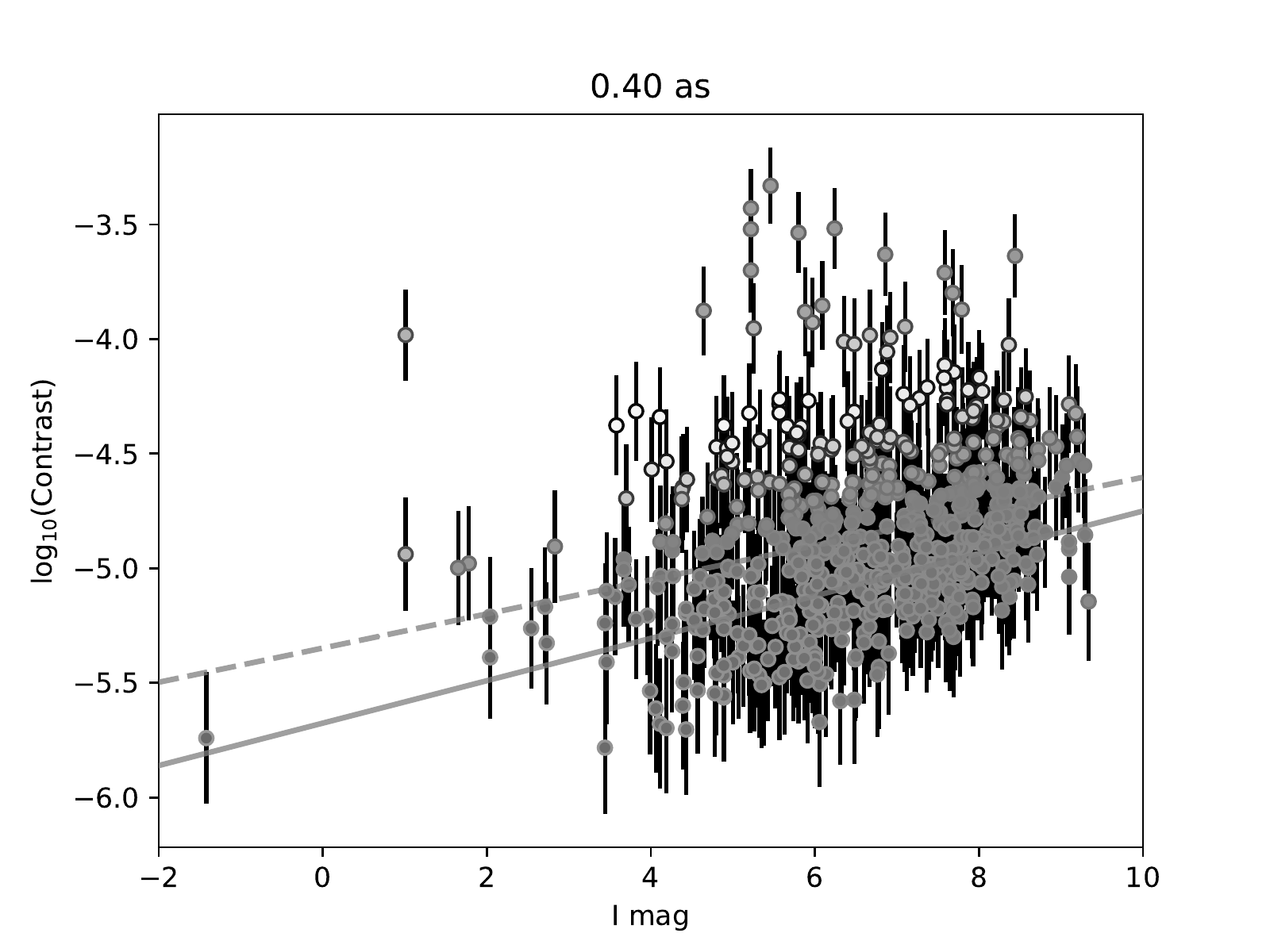}
\caption[]{0.4 arcseconds}
\end{subfigure}
\begin{subfigure}[c]{0.495\textwidth}
\includegraphics[width=\columnwidth,clip=true,trim = 0.1in 0.1in 0.5in 0.25in]{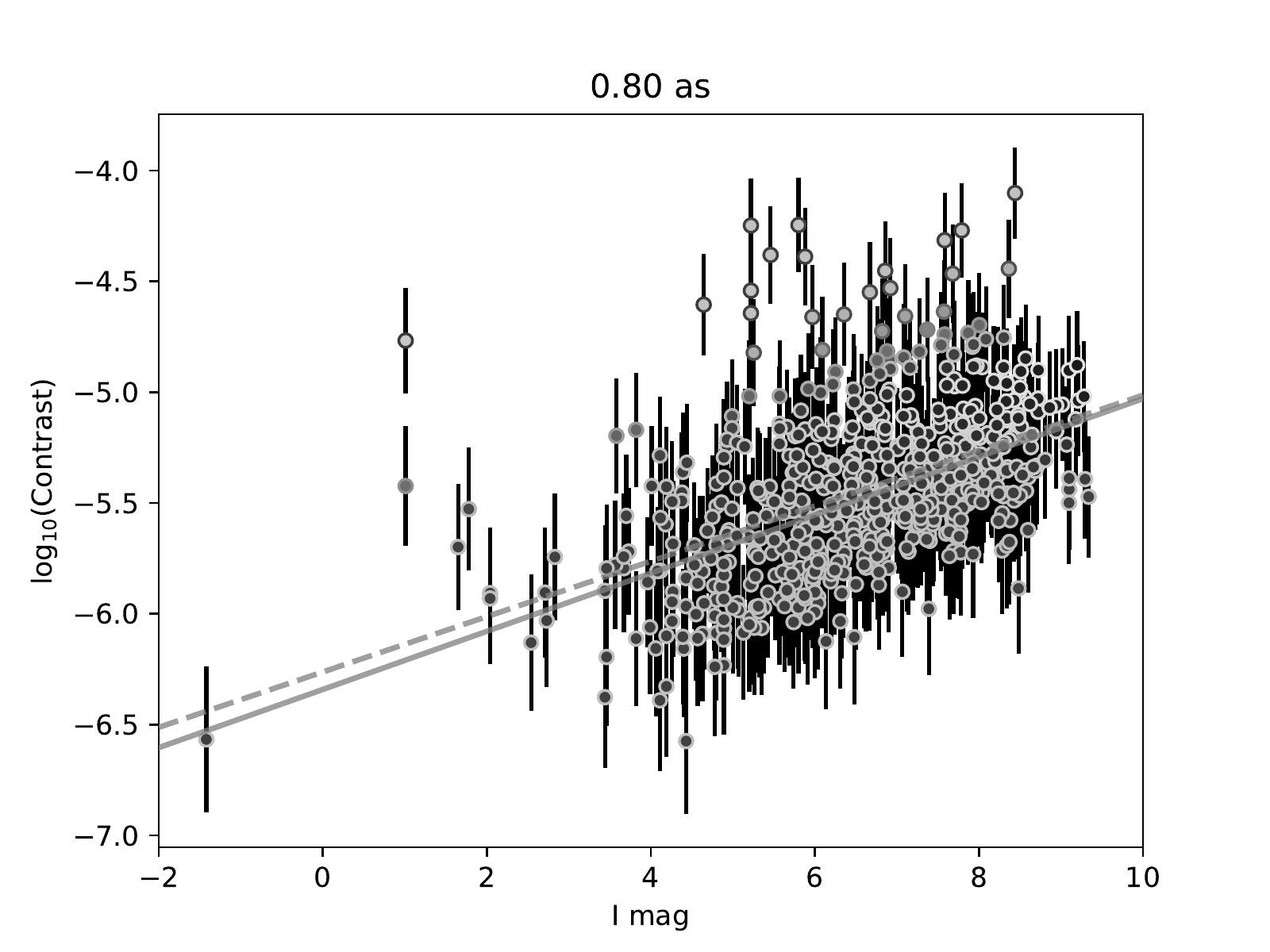}
\caption[]{0.8 arcseconds}
\end{subfigure}
\begin{subfigure}[c]{0.495\textwidth}
\includegraphics[width=\columnwidth,clip=true,trim = 0.1in 0.1in 0.5in 0.25in]{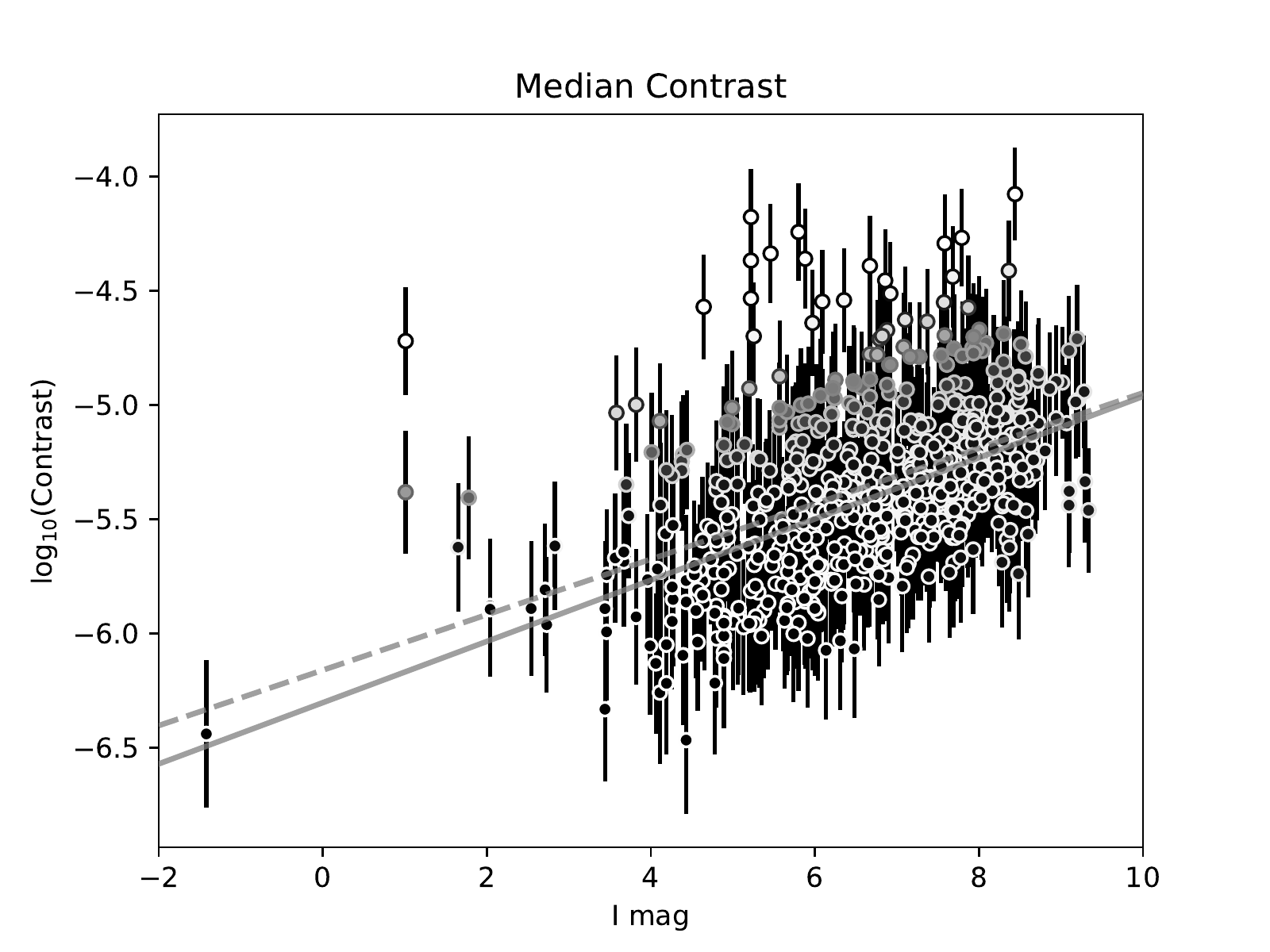}
\caption[]{Median contrast}
\end{subfigure}
\caption{Linear models of $\log_{10}$ contrast evaluated at different angular separations and the full curve median value as a function of target I band magnitude.  Points are shaded according to their probability of belonging to the foreground data set (black meaning unity probability and white meaning zero).  The dashed line shows the linear fit using all data, and the solid line shows the fit using only points with $>0.5$ probability of being in the foreground set. \label{fig:imagfits}}
\end{figure}

\begin{figure}[ht]
\centering
\begin{subfigure}[c]{0.495\textwidth}
\includegraphics[width=\columnwidth,clip=true,trim = 0.1in 0.1in 0.5in 0.25in]{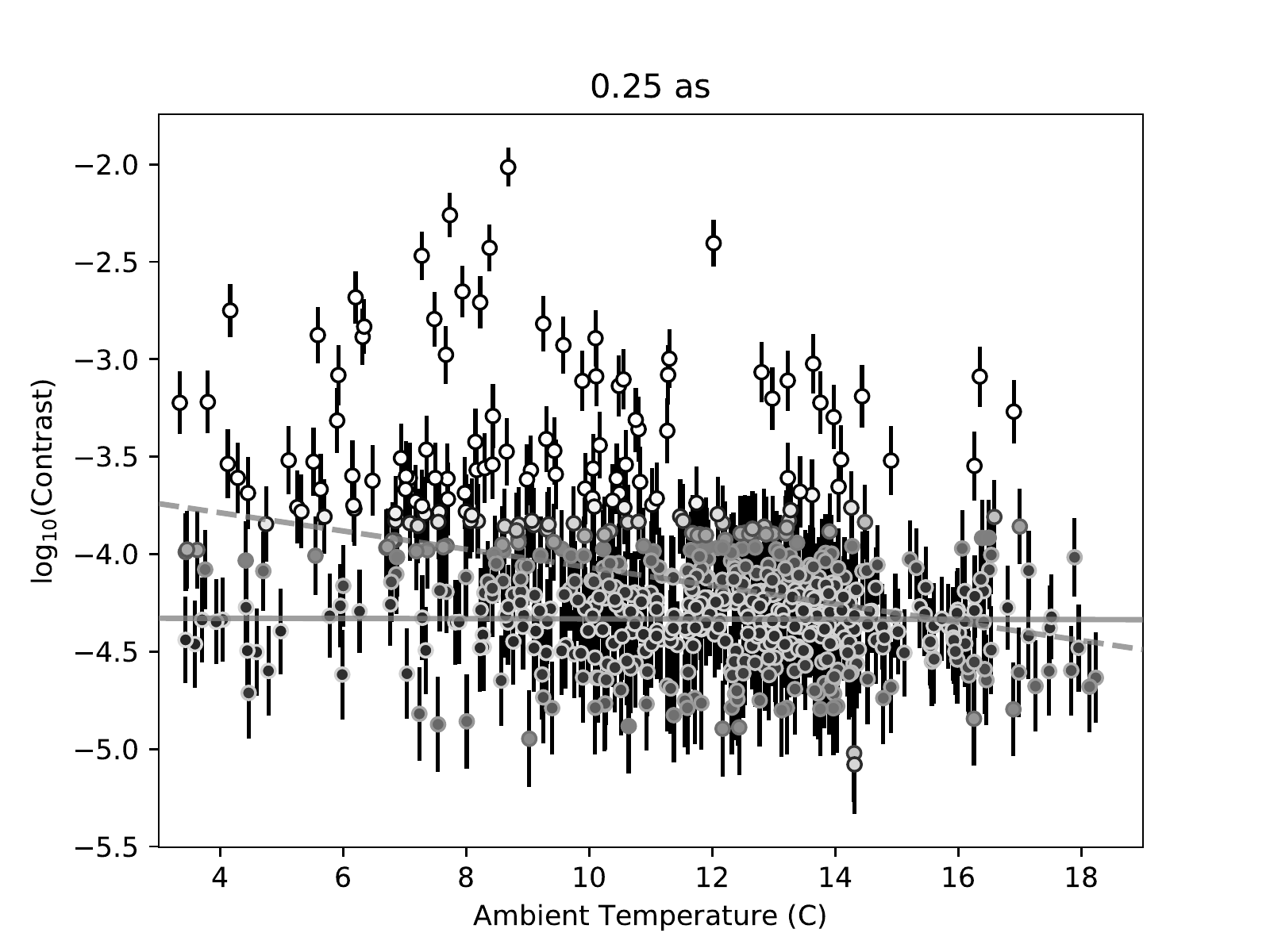}
\caption[]{0.25 arcseconds}
\end{subfigure}
\begin{subfigure}[c]{0.495\textwidth}
\includegraphics[width=\columnwidth,clip=true,trim = 0.1in 0.1in 0.5in 0.25in]{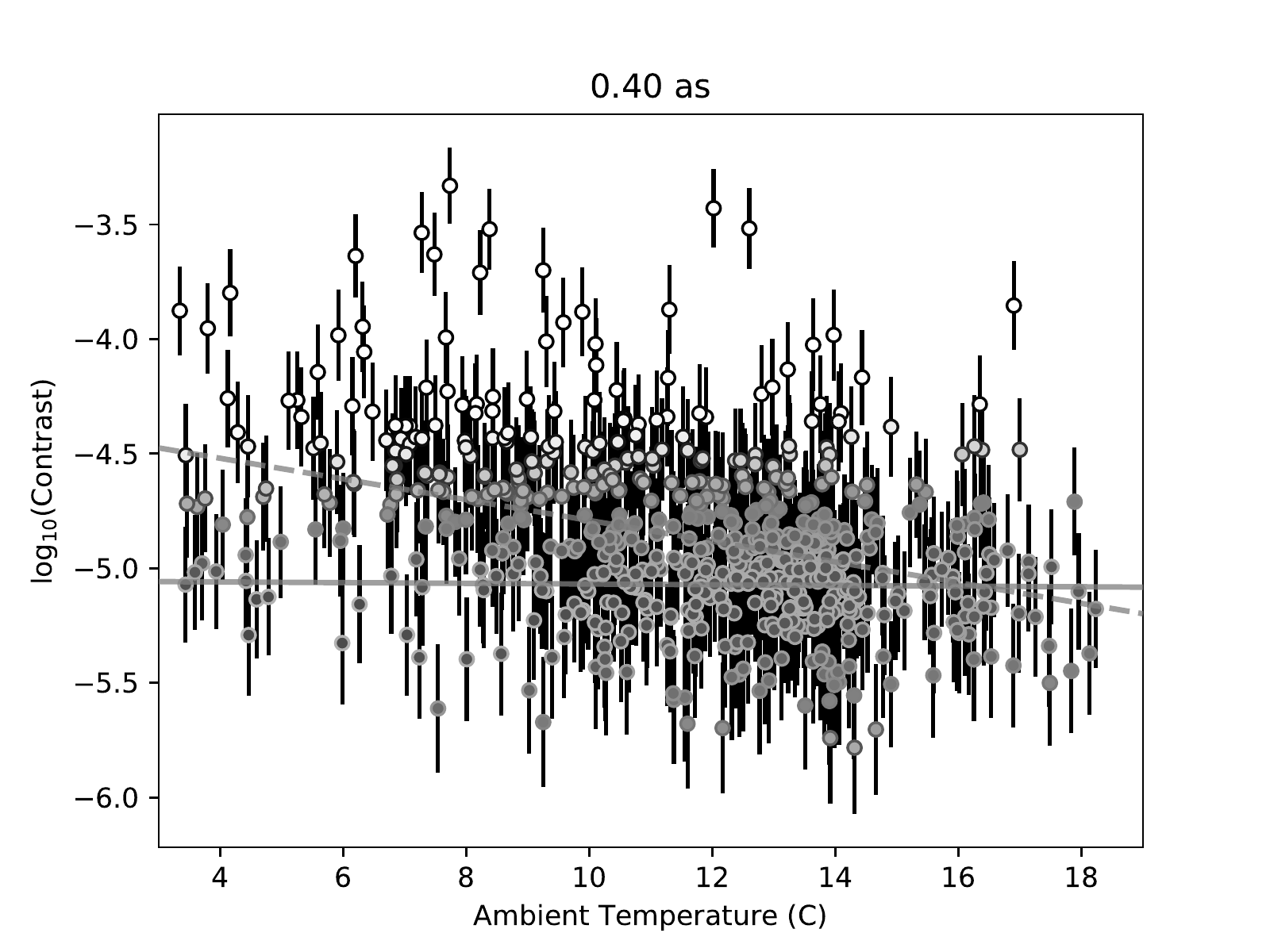}
\caption[]{0.4 arcseconds}
\end{subfigure}
\begin{subfigure}[c]{0.495\textwidth}
\includegraphics[width=\columnwidth,clip=true,trim = 0.1in 0.1in 0.5in 0.25in]{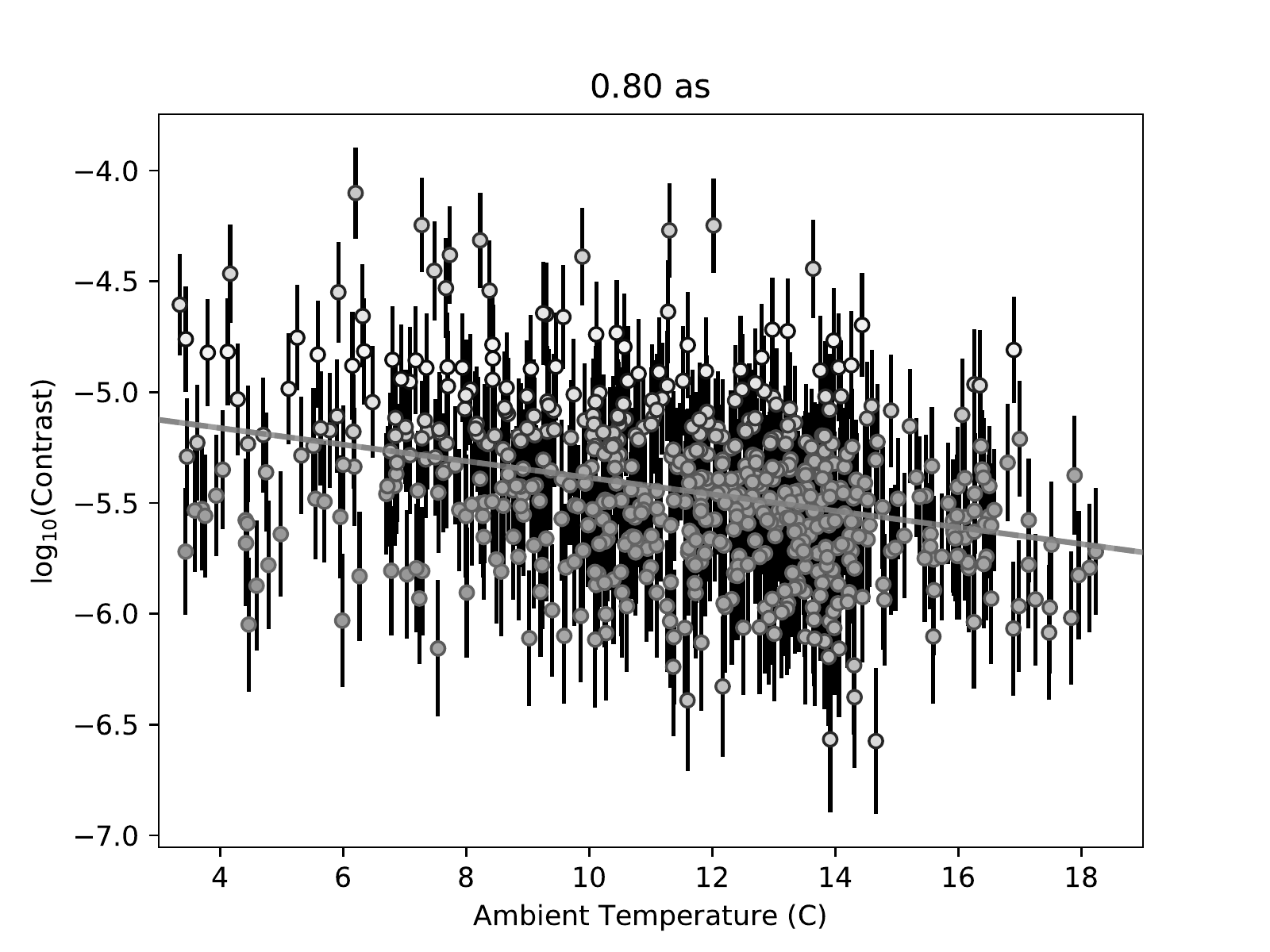}
\caption[]{0.8 arcseconds}
\end{subfigure}
\begin{subfigure}[c]{0.495\textwidth}
\includegraphics[width=\columnwidth,clip=true,trim = 0.1in 0.1in 0.5in 0.25in]{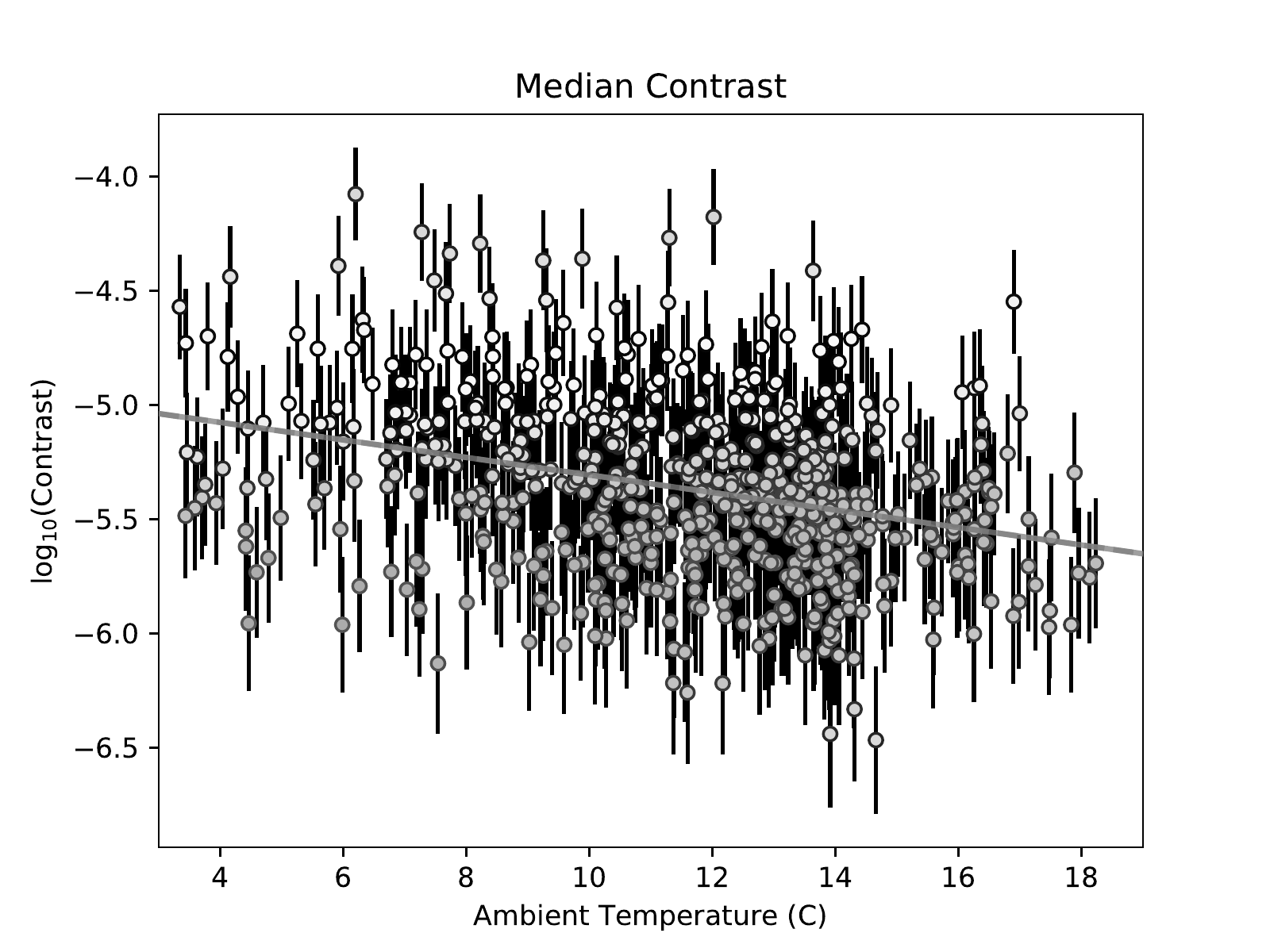}
\caption[]{Median contrast}
\end{subfigure}
\caption{Linear models of $\log_{10}$ contrast evaluated at different angular separations and the full curve median value as a function of mean ambient temperature during the observing sequence.  Color and line styles are as in \reffig{fig:imagfits}. In the case of the 0.8 arcsecond and median value plots, the dashed and solid lines fall on top of one another. \label{fig:ambtempfits}}
\end{figure}

Given the likelihood and the model, we can infer the model parameters, including the moments of the outlier distribution, via Markov Chain Monte Carlo. For the contrast variance, we use values derived from injection tests using synthetic planets added to real data to establish the errors on the derived contrast values\cite{ruffio2017improving,pueyo2016detection}.  We use the open source package \verb+emcee+\cite{foreman2013emcee} to carry out the sampling.

Figures \ref{fig:imagfits} and \ref{fig:ambtempfits} show the results of carrying out this fitting procedure using linear models of contrast as a function of the target I band magnitude and the mean ambient temperature during the observing sequence, respectively. The figures show both the linear fits derived from the full data set, as well as those using only the points which have greater than 0.5 probability of belonging to the foreground set.  In both cases, there is significantly more deviation between these fits for the contrasts evaluated at 0.25 and 0.4 arcseconds than the 0.8 arcsecond and median contrasts.  Most interestingly, in the case of the ambient temperature models, the linear dependence is entirely removed at 0.25 and 0.4 arcseconds by partitioning the data into foreground and outliers, whereas the 0.8 arcsecond and median contrast fits are not impacted in any significant way by outlier classification.   No higher order polynomial model can match the performance (in terms of RMSE and model convergence) of the linear model.

While these results are interesting, and hint at different physical processes governing image noise and contrast at varying angular separations (see \refnum{poyneer2016performance}), these models are not suitable for the prediction of contrasts from future observations. While the fits to the foreground data produce better RMSE values (at the levels of 0.4-0.5 $\log_{10}$ contrast) than the fits to the full data set, this methodology does not provide us with the means to determine whether a given observing sequence will fall into the foreground set.   Furthermore, both univariate and multivariate models of the data available prior to the start of an observation (the target properties) are highly noisy and do not produce reliable contrast predictions.

\section{REGRESSION NEURAL NETWORKS}\label{sec:dnns}

An alternate approach to multivariate modeling is to use artificial neural networks for data regression\cite{specht1991general}.  Many variations exist in the literature, but the basic regression network is a deep (multiple hidden layer) neural net with an input layer composed of continuous or categorical data, and a single output node trained to produce a continuous numerical value.  Training can be done via standard stochastic gradient descent, using sigmoid or atanh activation functions for the intermediate nodes. The great utility of this approach is that it does not require any model at all, and can, in theory, adapt to highly variable data without having to make specific assumptions about outlier distributions or even an explicit statement of the noise characteristics of the data.

\begin{figure}[ht]
\centering
\includegraphics[width=0.6\textwidth,clip=true,trim = 0.1in 0.1in 0.5in 0.25in]{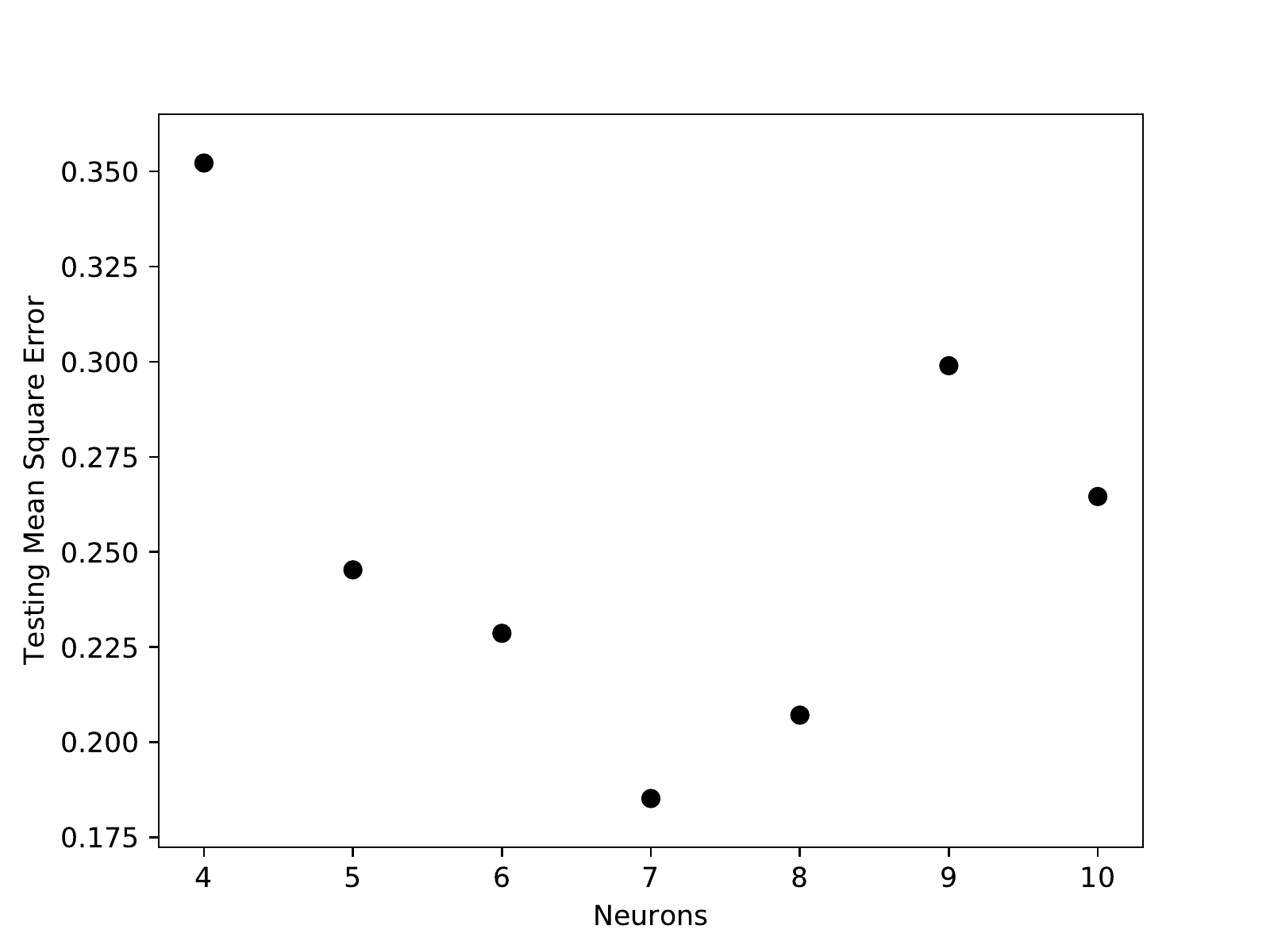}
\caption{RMSE values for 9 input layer regression DNNs with a variable sized, single hidden layer.\label{fig:single_layer_errors}}
\end{figure}

Here, we present a series of regression networks, trained on various partitioning of the available data and scored for their predictive power.  We use the open source TensorFlow software\cite{abadi2016tensorflow}, taking advantage of the code's semi-automated network building utilities to construct our models.  In each case, we partition the data into a training set, composed of 75\% of the data, selected at random, and a testing set composed of the remaining 25\%.  A primary drawback of regression networks is the ease with which they can be overfit to a particular training set.  This is especially true in the cases where the input layer is small, as compared with the hidden layers, and a limited amount of training data is available. To evaluate the potential for overfitting of our data, we carried out numerical experiments with single hidden layer networks with varying numbers of nodes, using a fixed number of inputs.  \reffig{fig:single_layer_errors} shows the results of varying the number of nodes in a single hidden layer for a 9-node input layer.  As the size of the hidden layer approaches the size of the input layer, performance degrades due to overfitting of the training set.

\begin{figure}[ht]
\centering
\begin{subfigure}[b]{0.395\textwidth}
\includegraphics[width=\columnwidth]{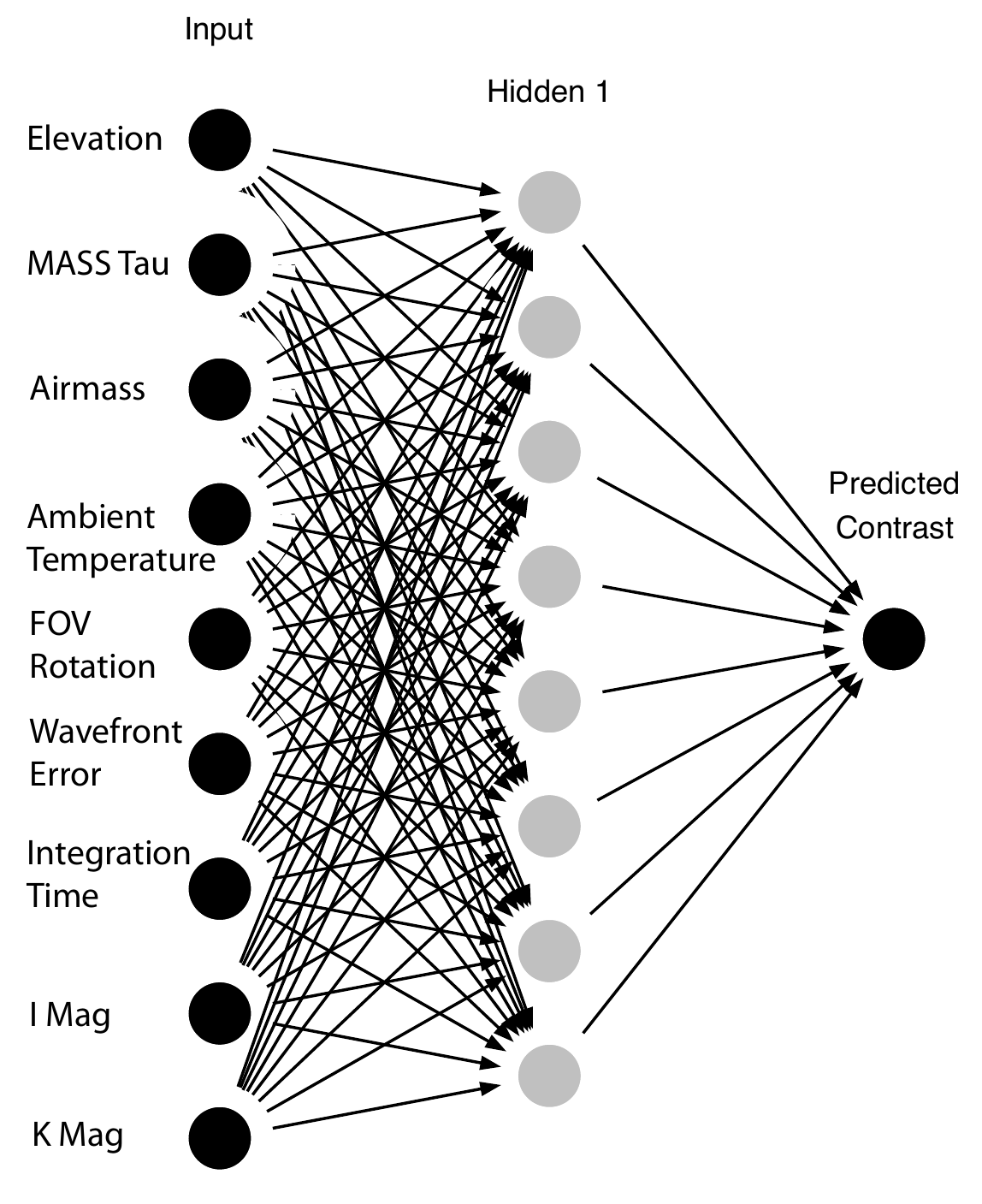}
\caption[]{Network 1 Architecture}
\end{subfigure}
\hfill
\begin{subfigure}[b]{0.59\textwidth}
\includegraphics[width=\columnwidth]{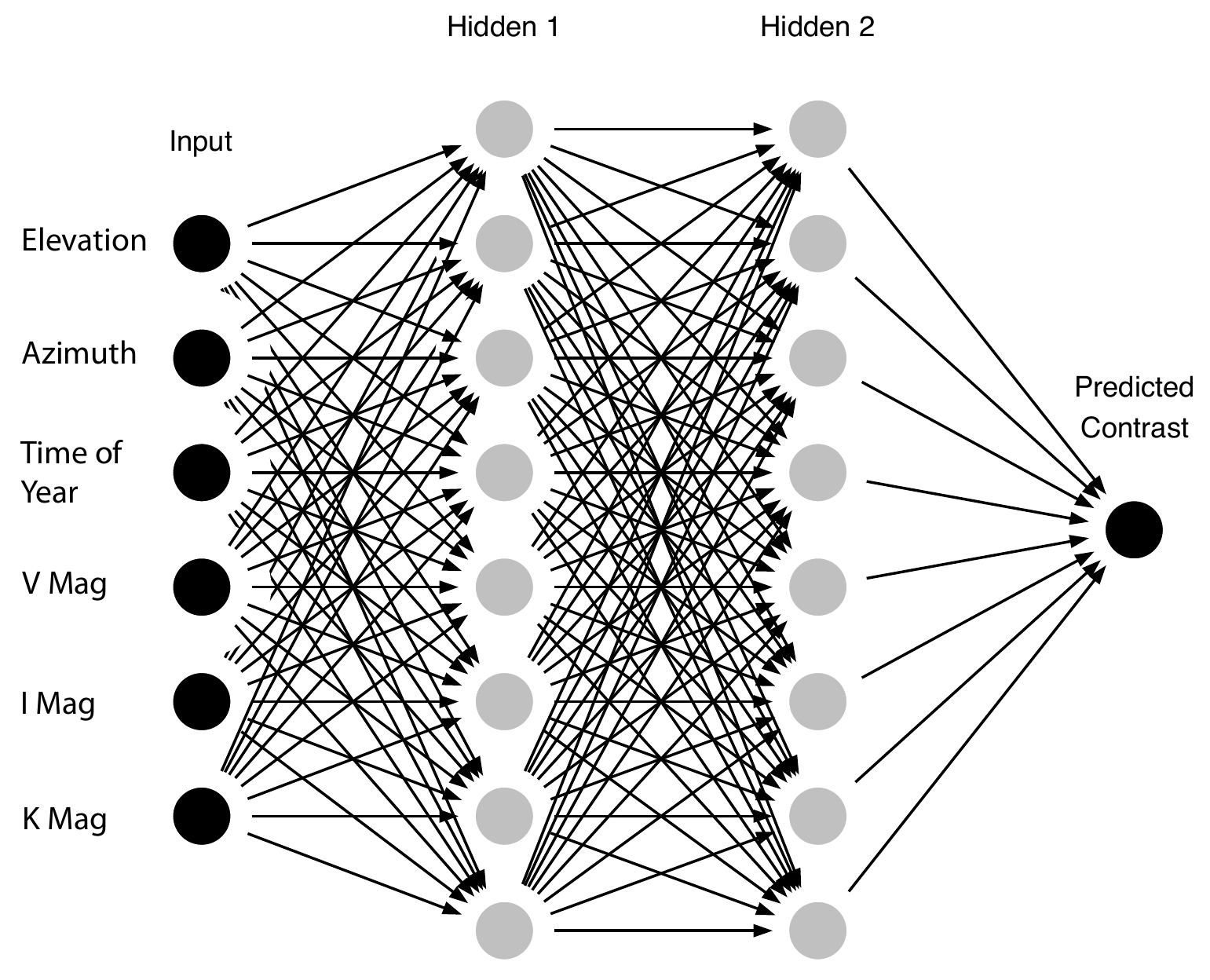}
\caption[]{Network 2 Architecture}
\end{subfigure}
\begin{subfigure}[c]{0.495\textwidth}
\includegraphics[width=\textwidth,clip=true,trim = 0.1in 0.1in 0.5in 0.25in]{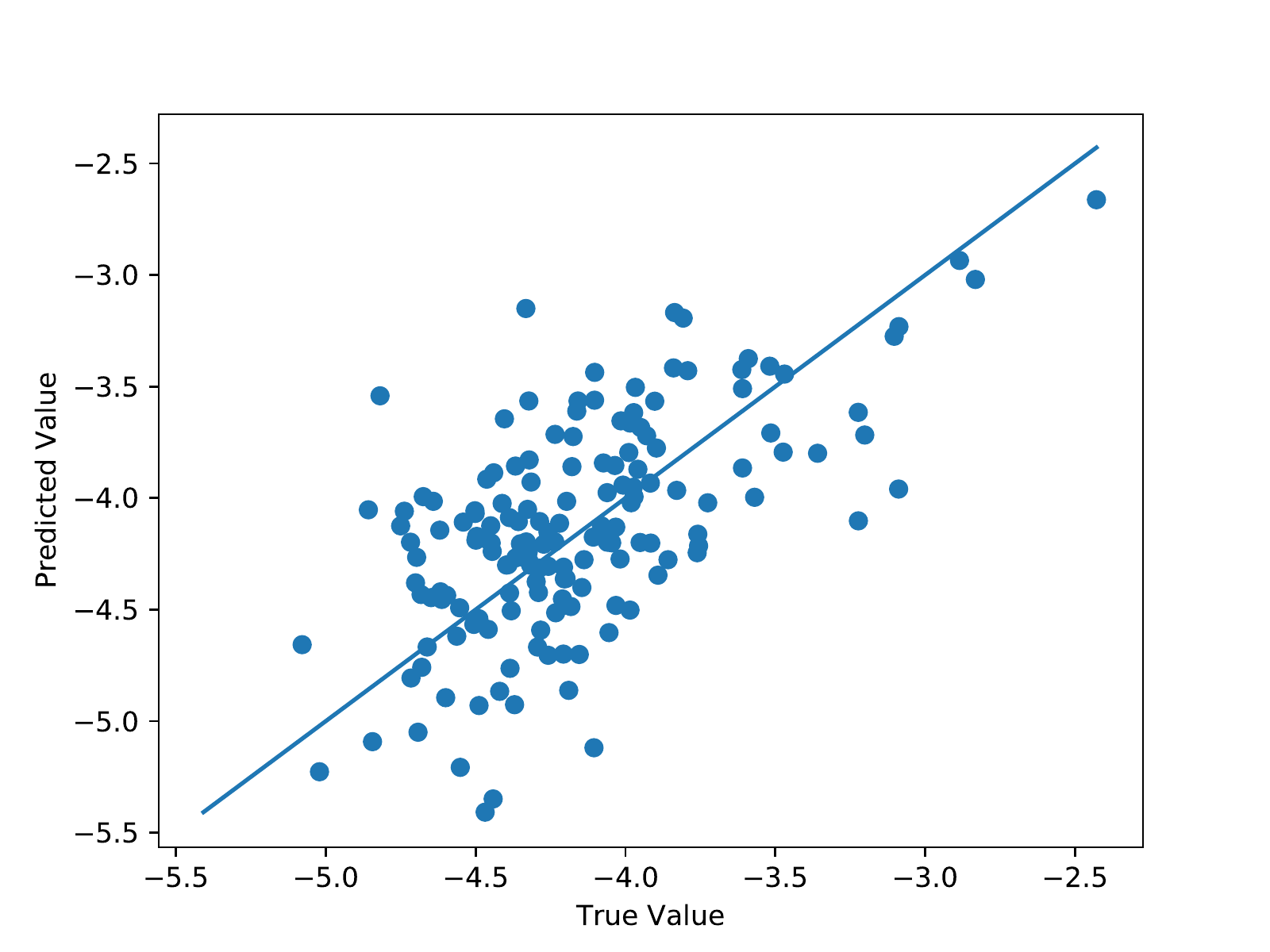}
\caption[]{Network 1 Test Result}
\end{subfigure}
\begin{subfigure}[c]{0.495\textwidth}
\includegraphics[width=\textwidth,clip=true,trim = 0in 0.1in 0.4in 0.25in]{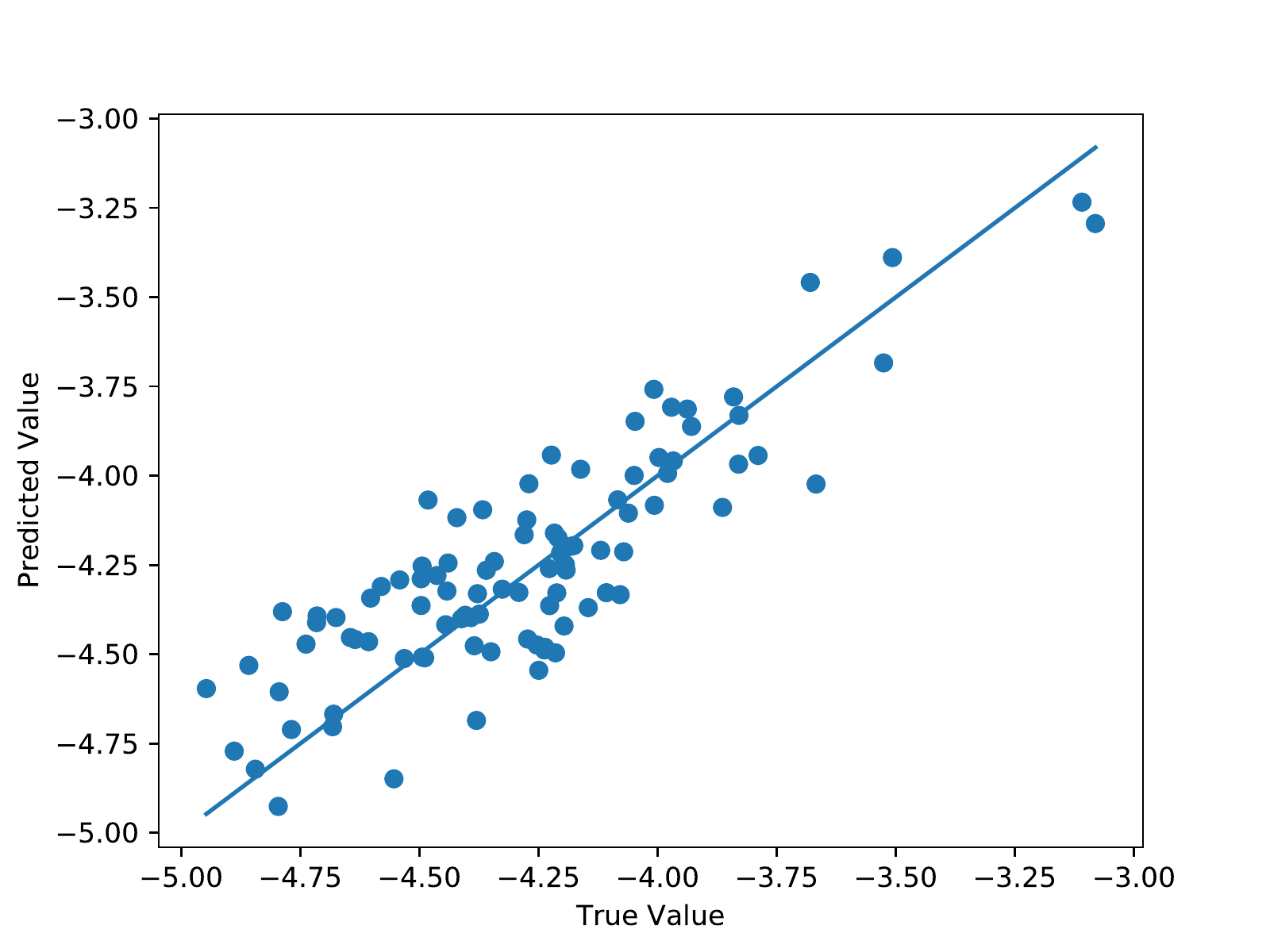}
\caption[]{Network 2 Test Results}
\end{subfigure}
\caption{Results from two regression network architectures. The left column shows a single hidden layer network with 8 nodes, trained using the 9 most highly correlated values from \reffig{fig:corr025}. The network achieves an RMSE of 0.40 $\log_{10}$ contrast on the testing set.  The right column shows a two layer hidden network with sixteen total hidden nodes, trained using 6 input values that would be available prior to the start of an observing sequence, and could be used for planning purposes.  The network achieves an RMSE of 0.18 on its testing set.  Both networks are fully connected and are trained for a fixed number of steps. \label{fig:dnnres1}}
\end{figure}

\reffig{fig:dnnres1} shows the architectures and testing results for two of the more successful regression network evaluated.  The first uses nine inputs, drawn from the most highly correlated values with contrast discussed in \refsec{sec:correlations}, which include both target properties and ambient and operating conditions.  The network has a single, fully connected, 8-node hidden layer and achieves an RMSE of 0.4 on its training set.  In general, single hidden layer networks can explain any linear relationships between the inputs and outputs, and so it is unsurprising that this network performs at the approximate level of the best linear fits from \refsec{sec:models}.

We next consider a network trained on only the six inputs available prior to the start of the observing sequence, and not including any ambient or operating conditions.  Here, the strictly linear correlations are weaker than in the nine input case, and so we explore a fully connected, two hidden layer network, with 16 total hidden nodes, which can, in principle, explain any categorical relationships that exist in the data.  After 10,000 training iterations, the network achieves an RMSE of 0.18, outperforming any linear (or higher-order polynomial) models from \refsec{sec:models}.  Given the input values and achieved error, this network could be used for long-term survey planning purposes if a contrast maximization strategy for specific targets were to be adopted (the GPIES survey seeks to create an unbiased sample via its target selection\cite{mcbride2011experimental,savransky2013campaign}, but other operating models exist). 

\begin{figure}[ht]
\centering
\includegraphics[width=0.8\textwidth,clip=true,trim = 0in 0.1in 0.4in 0.5in]{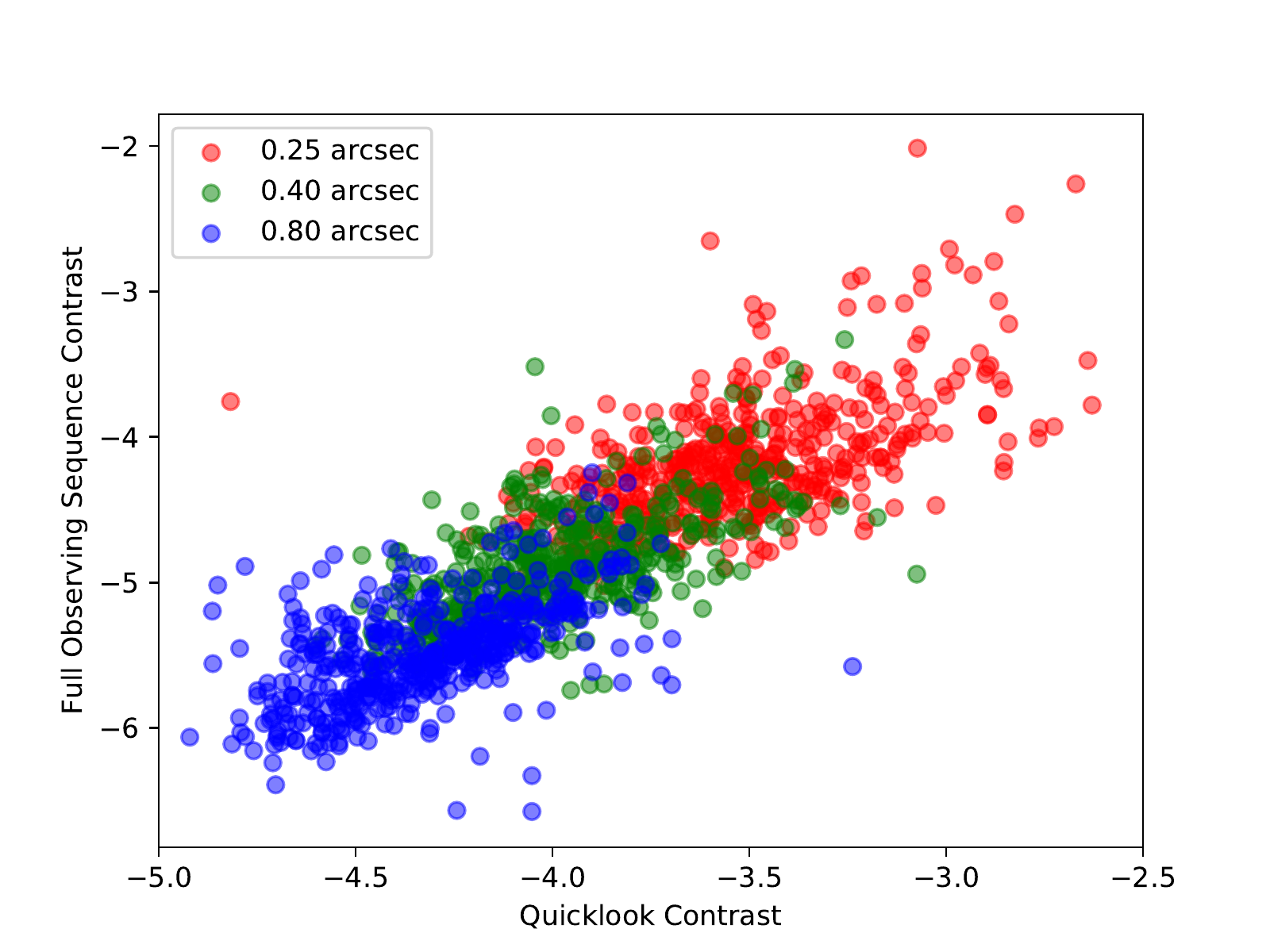}
\caption{Scatter plot of quicklook contrasts at 0.25, 0.4 and 0.8 arcseconds from the first image in an observing sequence versus interpolated values at the same separations from the final post-processed full sequence contrast.\label{fig:quicklookcontrasts}}
\end{figure}

We can also consider potential predictive capabilities given all of the information available at the very start of an observing sequence (i.e., after the completion of the first one minute exposure in an hour-long sequence).  This type of modeling is highly applicable to queue scheduled surveys with absolute guarantees not just on conditions, but on overall science data quality.  In addition to all of the parameters in Figures \ref{fig:corr025}-\ref{fig:corr080}, in this representing only the metadata of the first image rather than the sequence average, we also have the quicklook contrasts from the initial on-summit quick reduction of the first image. \reffig{fig:quicklookcontrasts} shows the quicklook-derived contrasts for the first image in an observing sequence at 0.25, 0.4 and 0.8 arcseconds plotted as a function of the final sequence contrast.  As with previously considered parameters, there is a generally linear relationship between these contrasts, with high scatter and multiple evident outliers.

\begin{figure}[ht]
\centering
\includegraphics[width=0.8\textwidth,clip=true,trim = 0in 0.1in 0.4in 0.25in]{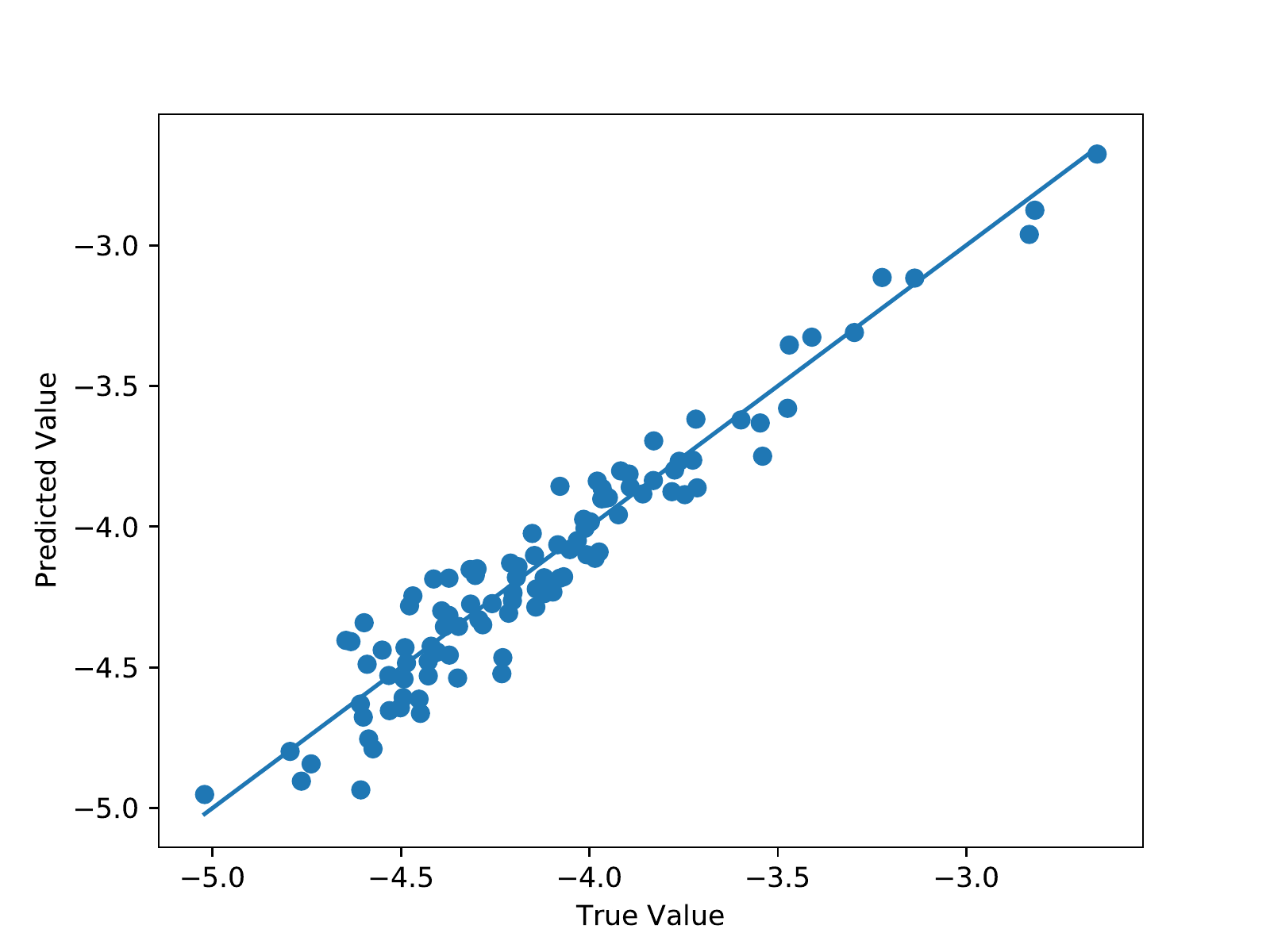}
\caption{Testing results from a 3 hidden layer network with 22 input nodes, 60 hidden nodes and a single output.  The achieved RMSE on the testing set is 0.11 $\log_{10}$ contrast. \label{fig:dnnres2}}
\end{figure}

We create a new regression network with 22 inputs, including all of the parameters from \reffig{fig:corr025} except for the total field of view rotation and total integration time, and representing only the first image of the sequence, along with the three quicklook contrasts from that image.  We could also include the full integration time and field of view rotation values under the assumption that the sequence will be completed as planned, but as all GPIES sequences are effectively the same in terms of integration time, and the total rotation is deterministically derivable from other target parameters, these two inputs end up being fully redundant under the assumption of fully executed sequences.  The network has three hidden layers of 20 nodes each and is trained with the same partitioned data set as the previous examples for a fixed 10,000 iterations. \reffig{fig:dnnres2} shows the testing results of this network, which achieves an RMSE of 0.11 $\log_{10}$ contrast, outperforming all previously considered models.  For comparison, the best linear fit to the full sequence contrasts as a function of the quicklook contrasts has an RMSE of 0.54  $\log_{10}$.  The practical application of such a model would be to give observers real time information on the expected outcome of completing the execution of a sequence.  In a time-limited survey, it may be highly beneficial to abort observing sequences that may produce poor results, especially if operating in queue mode with condition and contrast guarantees. The scatter of the model (at least for the testing set considered) is sufficiently small to make false positives unlikely, giving the model good predictive power. 

\section{CONCLUSIONS}\label{sec:conclusion}

We have presented a variety of analyses of GPIES contrast data as a function of target properties, observing conditions and instrument operating conditions, using methodologies from correlation analyses, Bayesian hierarchical networks, and machine learning.  All of the models considered were driven directly by the data, with no consideration for underlying physical processes or any intuition into the specific operation or performance of the instrument.  While some of the predictive models shown here are promising, significant additional validation would be required before they could be used for planning and decision making.

In particular, the data set used represents four years of survey operations, during which time changes were made to the instrument (in particular software updates and fixes to mechanical subsystems), and various minor changes were introduced into operating procedures and observation scheduling rules as the team learned more about the particulars of observing at Gemini South.  The basic observing sequence has, however, remained at 40 images in H band over the course of one hour, which has allowed us to perform the analyses described here.  The models do not contain any factor that accounts for temporal variation of performance, other than the day of year number, which is primarily intended to account for seasonal variations.  As such, the models cannot capture any potential changes in overall survey performance over time, and thus may not be explicitly applicable to future observations.  Similar concerns would apply to any models of this type trained for other surveys.  Practically, if such modeling were to be included in survey planning, the models would need to be consistently revalidated as new data is gathered, to ensure that any model drift is captured.  At the same time, an important unanswered question is the minimum amount of data that must be collected in order to produce reasonable performance models.  We can start to answer this question now by further partitioning the training data set, which we will reserve for future work.

\clearpage 

\acknowledgments
This research made use of Astropy, a community-developed core Python package for Astronomy (Astropy Collaboration, 2013), Matplotlib\cite{hunter2007matplotlib}, Pandas\cite{mckinney2010data} and TensorFlow, an open source software library for high performance numerical computation developed and released by Google Inc. with community support.

\bibliography{Main}   
\bibliographystyle{spiebib}

\end{document}